# P-type doping of elemental bismuth with indium, gallium and tin: a novel doping mechanism in solids

Hyungyu Jin[1], Bartlomiej Wiendlocha[1,2], and Joseph P. Heremans[1,3,4]

1. Department of Mechanical and Aerospace Engineering, The Ohio State University, Columbus, OH, USA
2. Faculty of Physics and Applied Computer Science, AGH University of Science and Technology, Al. Mickiewicza 30, 30-059, Krakow, Poland
3. Department of Physics, The Ohio State University, Columbus, OH, USA
4. Department of Materials Science and Engineering, The Ohio State University, Columbus, OH, USA

**ABSTRACT**

A new doping mechanism is described, whereby a doping impurity does not simply transfer charge to the bands of a host semiconductor or semimetal, but rearranges the core energy levels deep in the valence band of the host. This, in turn, leads to a redistribution of all electrons in the host, and, if designed properly, changes the location of the Fermi level $E_F$ and the density of conducting charge carriers near $E_F$. The principle is proven experimentally in elemental Bi, whereby group III elements In and Ga dope Bi p-type, in spite of the fact that all three atoms are trivalent in the solid state. Electronic structure calculations show the formation of a hyperdeep defect state (HDS) and its effect on the $E_F$ in Bi doped with In (Bi:In) and Ga (Bi:Ga). The HDS at ∼ 5 to 6 eV below the $E_F$ of elemental Bi hybridizes with the Bi $6p$ electrons, and deprives the Bi valence band of two electrons per impurity atom. This then perturbs the electron count in the solid and lowers the $E_F$. The charge on the impurity atoms is unchanged. In principle, this doping action does not result in the appearance of ionized impurities that scatter conduction electrons and holes in conventional doping. Experimentally, Shubnikov – de Haas and Hall Effect measurements show that adding In to Bi results in an increase in the density of holes. Thermoelectric, galvanomagnetic and thermomagnetic data are given for single-crystal and polycrystalline samples. In-doping leads to an enhancement of the thermoelectric figure of merit, which suggests that the new doping mechanism also gives a route to develop better thermoelectric materials. The same mechanism is at work for Sn-doped Bi (Bi:Sn), although here ionized impurity scattering is not avoided.

# I. INTRODUCTION

Controlling and engineering the concentration of free charge carriers is the mechanism by which functionality is created in all solid materials used for the generation of renewable energy, including solar cells[1], thermoelectric semiconductors[2], and photoelectrochemical cells[3] designed for water splitting. Indeed, this capability, called doping, is crucial to most of electronic functions such as rectification, amplification and gating. Its mastery was a necessary condition for semiconductors to displace vacuum tubes and make modern electronics possible. Doping enables the engineering of local variations in the equilibrium charge carrier concentrations. This in turn enables the design of semiconductor structures in which the current/voltage relations are not linear, a necessary condition in the design of active electronics. The non-linear current/voltage dependences also create the potential differences in solar or photoelectrochemical cells. Even in linear devices, such as thermoelectric Seebeck generators and Peltier coolers, doping is necessary because the electron or hole concentrations must be optimized[2] to create a high thermoelectric figure of merit $zT$.[4]

Doping is achieved either by controlling the native defects, or by chemical substitution of atoms of the host lattice with extraneous "impurity" atoms. In the case of defect-doping, one uses either vacancies or anti-sites in the materials lattice. In the case of vacancies, a specific lattice site is devoid of an atom which otherwise would have a charge. Anti-site defects in compound semiconductors of formula AB, where an A atom sits on a lattice site where a B atom would normally be found, alter the free carrier concentration when there is a charge difference between the two atoms in the lattice. Impurity doping means that extraneous atoms are deliberately added (which in the example above would be neither A nor B) that alter the free electron concentration. These extrinsic impurity atoms can either transfer electrons to the conduction band of the host solid, and act as electron donors; or they can capture charges from the valence band of the host, leaving holes behind, and act as acceptors. In this article we describe a new way by which extrinsic atoms can achieve this task. The simplest and classical view of how these donors and acceptors work in semiconductors is the rigid-band approximation: the assumption that the band structure of the host solid remains unchanged by the presence of a dilute amount of impurity atoms. The sole role of the impurity atoms is then to transfer charge to the host, leading to a change in the number of free carriers in the solid. This only works if the effective valence of the impurity atoms differs from that of the atoms in the host. In this model, the transfer of charge from impurity atom to host leaves the impurity atom ionized, which, in turn, scatters the free electrons and can lower their mobility. We show here that one can induce doping in an elemental solid even with dilute amounts of an isovalent atom, leading to a departure from the rigid-band approximation, while leaving the impurity atom without a net charge.

This discovery can lead to new concepts in semiconductor-based technologies. In principle, if the impurities are not ionized in the conventional sense by this new doping mechanism, the free electrons are less scattered and retain higher mobility. This may lead to better performing devices than when using conventional doping mechanisms. The advance this may represent is more likely to find its way first in semiconductors used in energy conversion: these disciplines make use of a large variety of new materials, for which dopants must be found.

Here, the principle of the new approach is demonstrated in bulk single crystals of elemental Bi (data on polycrystals are in the Supplemental Online Material, SOM), not because that is a technologically relevant material, but because it provides an experimental platform that avoids possible parasitic effects, particularly defect-doping which can be hard to reproduce from sample to sample. Being an element, Bi cannot host anti-site defects, and Bi is not known to harbor vacancies. Elemental Bi is a semimetal on which many physical properties have been discovered, from the Seebeck effect[5] itself to some of the most timely topics in solid-state physics.[6,7] This is due to several advantageous features of Bi: its unique electronic structure; low electron concentration; extremely small electron mass and accompanying high electron mobility; high atomic number, the highest (83) of all non-radioactive elements, which leads to strong spin-orbit interactions; and its sole naturally occurring isotope (209) with a $10^{19}$ years lifetime that avoids isotope scattering. The band structure of Bi was investigated experimentally.[8,9,10,11,12,13] The simultaneous presence (Fig. 1) of an electron band, crossing the Fermi level $E_F$ around the L-point of the Brillouin zone, and a hole band, centered at the T-point, makes Bi a semimetal with a small band overlap (~ 37 meV at 4.2 K) and a small and equal number of electrons and holes (~ $3 \times 10^{17}$ cm$^{-3}$ at 4.2 K, ~ $3 \times 10^{18}$ cm$^{-3}$ at 300 K[14]).

The transport properties of Bi are reviewed by Issi[14], yet new effects are still being discovered.[15] The simultaneous presence of electrons and holes results in a compensation between their partial thermopowers, and thus, a limited net thermopower and $zT$. If the energy overlap between conduction and valence bands were somehow lifted, i.e. if elemental Bi were a semiconductor, its $zT$ is calculated to reach 1.3 at room temperature.[16] However, this has never been achieved in elemental Bi. Bi$_{100-x}$Sb$_x$ alloys are isoelectronic with elemental Bi, with which they share very similar lattice and electronic band structures. Alloys with ~7% < $x$ < ~20% are semiconductors and are the most promising n-type thermoelectric materials below 200 K.[17,18,19] As with all thermoelectric materials, it is necessary to dope Bi$_{100-x}$Sb$_x$ alloys optimally in order to maximize $zT$. This is achieved with the same acceptors and donors as in Bi, the element this study focuses on for simplicity. It is likely that the principles developed here apply also to the tetradymites of general formula (Bi$_{1-x}$Sb$_x$)$_2$(Te$_{1-y}$Se$_y$)$_3$, which share the crystal structure of Bi and are technologically the most important thermoelectric semiconductors, used in Peltier cooling at room temperature.[2] Finally, because Te is not an earth-abundant element, active research on Te-less substitutes for the tetradymites is ongoing, and the Bi$_{100-x}$Sb$_x$ alloys are amongst the best n-type candidate materials.

Few electrically active impurity atoms are currently known for Bi and Bi$_{100-x}$Sb$_x$ alloys: Pb[17,20,21] and Sn[20,22,23] are acceptors, while Se,[20] Te,[22,24] and Li[25] are donors. Changes in the electrical resistivity $\rho$ were observed[20] in Bi doped with group III elements that are consistent with the work we present here; however, they were not understood at the time. Here, we add the isovalent atoms In (Bi:In) and Ga (Bi:Ga) to the list of acceptors, and describe the doping mechanism, which differs considerably from conventional charge-transfer doping.

A simple electron count (e.g. In$^{1+}$ in Bi$^{3+}$, or In$^{3+}$ in Bi$^{5+}$) can lead to the conclusion that group III elements can release two holes per atom. The band structure calculations reported here show that the reality is more subtle: the 6$s$-electrons of Bi contribute a density of states (DOS) deeper in the valence band block than the In 5$s$ electrons, leaving merely its three Bi 6$p$-electrons to determine the bands around $E_F$ (Fig. S1 in SOM); in this sense, the effective valence of Bi in the solid is three. Therefore, group III elements are not expected to transfer charge to the electronic bands of Bi, as each In or Ga atom possesses the same number (3) of outermost electrons as each Bi atom in the solid. The calculations then reveal that the In 5$s$- electrons hybridize with the Bi 6$p$-electrons in the valence band block in a special hyperdeep state, effectively immobilizing them and taking them out of the sea of conducting electrons near $E_F$. The charge transfer occurs not between delocalized electrons near $E_F$ and the doping impurity, but between Bi $p$-electrons far from $E_F$ and the In atom, which in turn rearranges the electrons in the whole solid and in particular decreases the number of electrons near $E_F$. This amounts to the In acting as an acceptor and constitutes a departure from the classical picture of impurity doping by simple charge transfer into rigid bands.

Hjalmarson et al.[26] and Hoang et al.[27] addressed non-rigid-band doping effects, and showed that certain impurity atoms in a semiconductor may form a pair of defect states: an electrically inactive "hyperdeep" defect state (HDS) lying below the valence band, and a "deep" defect state (DDS) located near the band gap. By analogy with molecular states, those defect states correspond to the bonding and anti-bonding pairs of orbitals, respectively. Examples are oxygen in GaAs[26] and In, Ga, and Tl in PbTe.[27] The DDS either can be hybridized with the host bands and located near $E_F$ where electrons in it conduct electricity, or located in the gap and act as a trap. The DDS is called a resonant state (or resonant level) when it is located in the main conduction or valence band block near $E_F$, creating an excess in DOS. Some resonant states (Tl in PbTe [28,29,30,31]; Sn in $Bi_2Te_3$[32]) distort the band structure of the host material in a way that leads to a significant increase in the thermopower over that which would be obtained at the same carrier concentration, resulting in enhancements of $zT$.[28,29,33]

We demonstrate that in Bi, the formation of the HDS from In 5$s$ or Ga 4$s$ and Bi 6$p$ states causes the p-type doping effect by depriving the Bi valence band of two electrons per In or Ga atom. The behavior of Sn in Bi is similar, with the Sn 5$s$ state forming a HDS with the Bi 6$p$ states. The fact that Sn is an acceptor was explained in the past by the simple electron count described above ($Sn^{4+}$ in $Bi^{5+}$), and assuming pure rigid-band-like behavior. But, it is already known[34] that this hypothesis cannot explain the temperature-dependence of the p-type doping of Bi. Here we show that the mechanism of p-type doping of Bi with Sn also involves a charge transfer from deep-lying Bi electrons to an HDS, resulting in a rearrangement of the total Bi electron population and a decrease in the concentration of delocalized electrons. The HDS is akin to that present in systems[32,33] where specific impurity dopants form resonant levels. The resonant level in those systems is actually the DDS near $E_F$ that accompanies the HDS, and there the transport properties of the host systems are not affected by the HDS, as it is located far away from $E_F$. In the case of In, Ga, and Sn in Bi, the DDS is simply absent, and the electrical activity on the host Bi comes from the HDS. This type of doping mechanism had not, to the best of our knowledge, been identified previously. Through the study of thermoelectric properties of Bi:In, we find that In also introduces energy-independent neutral impurity scattering, and not ionized impurity scattering. What the presence of the HDS modifies is the distribution of electrons in the total solid between those that are localized deep in the valence band and those near $E_F$ that participate in transport. Section II describes how the delocalized electrons see the same local charge on each In atom as they see on each neighboring Bi atom, and therefore are not subject to ionized impurity scattering. The $zT$ of Bi:In is much improved over those of undoped Bi and Bi:Sn with a similar carrier concentration. This behavior should be a general characteristic of isovalent doping in any solid system, and may well open a new route to improve the efficiency of thermoelectric materials.

This paper is structured as follows. Both experimental and theoretical results are presented on each of the three cases studied, Bi:In, Bi:Ga (given in SOM), and Bi:Sn. We present the theoretical band structure calculations first, starting with undoped Bi. Two complementary methods are used to treat the presence of dilute In, Ga, or Sn atoms: supercell calculations using the full potential linearized augmented plane wave method (FP-LAPW, WIEN2k code),[35] and calculations based on the Korringa-Kohn-Rostoker method, with the coherent potential approximation (KKR-CPA).[36,37,38] Both techniques independently verify the main results, yet each technique allows extensions that the other does not. The KKR-CPA proves the formation of HDS and the absence of a DDS, allows continuous observation of the shift in $E_F$ while impurity concentration is varied, and is used to calculate the effects of possible multiple defect structures. The supercell calculations reveal the lattice relaxation effects and the local bonding behavior around impurity atoms. The experimental section of the manuscript follows the theoretical one, with a section each on Bi:In (Bi:Ga in SOM) and Bi:Sn. For Bi:In, we report data on single crystals, using both Shubnikov-de Haas (SdH) and Hall effect measurements to show the p-type doping action of In. Thermoelectric and thermomagnetic measurements and the $zT$ are also given. For Bi:Ga, we present in the SOM $\rho$ and galvanomagnetic measurements on a polycrystalline sample, for which we develop a phenomenological model to extract mobilities and carrier densities. We also tried to dope Bi with aluminum (Al), but were not successful in adding a sufficient amount into the Bi lattice to observe a change in electrical properties of the material; if we accept that substitutional Al would also be an acceptor, this lets us put the solubility limit for Al in Bi to be < 5 ppm (atomic ratio). Lastly, because an extensive literature already exists[22,39] on Bi:Sn, only one single crystal Bi:Sn sample is studied that is doped with a similar hole concentration with the Bi:In samples. Thermoelectric properties and $zT$ on this sample are reported and compared with those on the Bi:In samples.

## II. THEORETICAL

Density Functional Theory (DFT) electronic structure calculations were performed, and all the technical details of the calculations are gathered in the SOM. Many papers model the electronic structure of undoped elemental Bi using semi-empirical pseudopotential or tight-binding calculations.[40,41,42] Two papers give first-principles calculations, namely pseudopotential calculations[43] and full-potential linear muffin-tin orbital (LMTO) calculations.[44] Our calculated electronic dispersion curves and DOS of Bi (full potential relativistic KKR[45] method, Fig. S1 in SOM) are in overall good agreement with experimental findings[9] and the previously published work.[43,44]

### II.1. Indium-doped bismuth

We start with the FP-LAPW supercell calculations for Bi:In. The supercell was constructed starting from the hexagonal equivalent of the Bi unit cell (containing 6 Bi atoms). This cell was multiplied, building a 4x4x1 supercell (containing 96 atoms, 1 In and 95 Bi, resulting in 1.04% ≈ 1% impurity atomic concentration), whose size is equivalent to 48 primitive A7 cells. The substitutional In atom creates a small negative chemical pressure, so that the three nearest neighbor (NN) and three next-nearest neighbor (NNN) Bi atoms move towards In. The distances are reduced from 3.06 Å to 3.03 Å (NN) and from 3.51 Å to 3.46 Å (NNN), respectively, in agreement with the ionic radii differences.

The total DOS divided by the number of atoms in the supercell is shown in Fig. 2(a); it is obtained from calculations for the relaxed system and includes the spin-orbit coupling. The salient feature is the resonant-like sharp peak in the DOS at approximately -5 eV, just below the main valence block. This corresponds to the HDS discussed in Refs [27,30] for group III impurities in PbTe. This peak is mainly formed by the 5$s$ states of In (Fig. 2(b)) and the 6$p$ states of Bi, and is not a simple semi-core-like 5$s$ state of In, as explained below. Some small amount of the In 5$s$ states also contribute to the $s$-like Bi DOS region centered at approximately -10 eV. Strikingly absent in the Bi:In system is the equivalent DDS near the $E_F$ (Fig. 2(c)), where we rather observe a broad and smooth In DOS maximum. This observation contradicts the general model of trap states in semiconductors,[26] or previous calculations for In or Tl doped PbTe,[29,30,31] where HDS follows DDS. The DDS is a resonant level in PbTe:Tl,[28,29,33] where it dominates transport, or a trap in PbTe:In.[27,30] In those two systems, the HDS was considered electrically inactive and not expected to affect the transport properties since its highly localized energy level is located far from $E_F$. In contrast, in Bi:In, it is the HDS that affects the transport properties because it activates a hole-like conductivity in the host system, as explained next.

Counting electrons, the HDS accommodates one electron from In 5$s$ state, and 1/6 of a Bi 6$p$ electron from each of 6 neighboring (3 NN and 3 NNN) Bi atoms for a total of one 6$p$ Bi electron per In atom. The HDS peak cannot be identified with the atomic-like semi-core state of the In atom, which would only accommodate both 5$s$ In electrons without modifying Bi states. To the contrary, by forming the HDS, each In atom effectively localizes one Bi electron, and indeed similar sharp localized peaks appear in the DOS of neighboring Bi atoms (Fig. S2 in SOM). Therefore, while one In atom substitutes one Bi atom without changing the effective valence (both being trivalent), the In substitution forces two electrons that otherwise would have been in the main valence band block to be confined into the highly localized HDS. This eventually results in an effect equivalent to creating two holes in the main valence band block. Thus, a single In atom, via the creation of the HDS, leaves two holes in the main valence band block. The resulting effect is that trivalent In behaves as an acceptor and moves $E_F$ of the system deeper in the valence band (Fig. 2(a)-(c)). Nevertheless, the number of electrons associated with the valence states of both elements, i.e. $5s^2 5p^1$ (In) and $6p^3$ (Bi), remains at three, confirming the electrically neutral character of In in Bi. Therefore, the In atoms should, in principle, not scatter conduction electrons or holes as strongly as ionized impurities would. This is because ionized impurity scattering involves Coulomb interactions between the delocalized electrons and the scattering centers, whereas neutral impurity scattering involves only the diffraction of the delocalized electron's wavefunction on the potential of the impurity atom; a much weaker interaction. This is a unique isovalent doping behavior that, to the best of our knowledge, has never been identified or observed before. The local binding feature of the HDS is visualized in the real-space charge distribution around the In impurity shown in Fig. 3. Here, the charge density corresponding to the sharp DOS peak at -5 eV is projected on the plane along the Bi-In bonds. This plane contains two NN and two NNN Bi atoms, as visualized in Fig. S3 of the SOM. The charge density shows $s$-$p$ hybridization, where the spherical $5s$ In electron cloud at the center is hybridized with the $6p$ Bi orbitals, creating the $s$-$p$ bonds. From the reciprocal-space point of view, this local bond would correspond to a dispersionless band, and the electronic states are expected to be localized.

To independently verify the results of the supercell LAPW calculations, and to study the effect of In concentration on DOS near $E_F$, KKR-CPA results are shown in Fig. 2(d) and Fig. 4. Consistent with the supercell calculations, KKR-CPA shows that substitutional In atoms form strongly peaked HDS-like DOS at about -5 eV below $E_F$. The difference is that the HDS DOS peak overlaps with the valence band DOS in the KKR-CPA results. The tentative explanation for the observed difference between supercells and CPA is related to the highly spatially localized nature of the HDS in the Bi:In system. In our example of a 96-atom supercell with 1 impurity atom, the HDS DOS peak comes almost exclusively from one In atom and 6 nearest Bi atoms. Thus, it may be viewed like a short-range order effect. In the CPA, which is an effective medium theory, the DOS is averaged over all possible configurations of atoms in the crystal. This results in the broadening of the peak, which in turn leads to the overlap because of the closeness in the positions of the peak and the valence band DOS. Despite this slight difference, the same conclusion that In introduces the formation of the HDS and dopes Bi p-type is still reached in the KKR-CPA results, as demonstrated below.

The KKR-CPA total and atom-decomposed DOS near $E_F$ for 0.1%, 0.5%, 1%, and 2% substitutional In in Bi are presented in Fig. 4, where contributions from In s- and p-states are also plotted. Confirming the supercell results (Fig. 2(c)), we observe that the partial In DOS has a broad maximum near $E_F$. This excess DOS consists of both s- and p-like contributions, and even for an In concentration as small as 0.1%, its full width at the half maximum (FWHM) is of the order of 500 meV. This is at least 10 times broader than the DOS peak introduced by the Tl $6s$ resonant state in the valence band of PbTe at the same concentration order.[33] Thus, absence of a resonant level near $E_F$ is confirmed, proving that the behavior of In in Bi is different from that of the group-III impurities in PbTe.[30] The main feature of the In electronic states near $E_F$ in Fig. 4 is the creation of the broad p- and s-like DOS 'hump' on the valence side of the pseudo-gap, predicted by both FP-LAPW and KKR-CPA methods. In practice, the DOS near $E_F$ is little affected by the small concentration of In. Increasing the In concentration moves $E_F$ to lower energies. Since the bands near $E_F$ are nearly rigid, a downward shift in $E_F$ corresponds to a decrease in electron concentration and increase in hole concentration, clearly showing a p-type doping behavior, which is impossible to explain in the absence of an HDS. The crystal field symmetry in the supercell calculations leads to a $p_z$ - $p_x+p_y$ splitting (along and perpendicular to the trigonal axis, the site symmetry is C3v), creating two main peaks in the In DOS. The spherical potential and CPA, which average electronic properties over all of the possible atomic configurations, smear those peaks into one p-like maximum, as can be seen in Fig. 4. As is a characteristic for disordered systems, smoothing of the DOS is observed when the In concentration is increased. Note that the partial In DOS is quite different from those of host Bi, demonstrating that In moderately modifies the electronic structure near $E_F$ in Bi beyond the rigid-band model.

We show in the SOM that Ga behaves in the same way as In in Bi. Moreover, in the next paragraph we will show that this mechanism is not even confined to only group III elements in Bi, thus is more general than we initially expected.

## II.2. Tin-doped bismuth

Sn has long been known as a monovalent acceptor impurity in elemental Bi.[20,22,34, 39, 46,47] At first glance, it seems intuitive that Sn, a group IV element, acts as an acceptor in Bi, a group V element through the simple charge transfer. However, we first point out that the effective valence of Bi is close to three, as explained in the introduction: in the rigid-band model, Sn could have been a monovalent donor ($Sn^{4+}$) as readily as a monovalent acceptor ($Sn^{2+}$). Second, an anomaly was observed[34] above 200 K in Sn-doped Bi: the acceptor action of Sn decreases rapidly with increasing temperature, a phenomenon that was attributed, without quantitative calculations, to the temperature dependence of the band structure. In that earlier model based on a simple rigid-band picture for dopants, the authors attributed this to a capture of electrons on an "acceptor level," which they realized must lie in one of the bands. They suggested that, if such an acceptor level exists, it must move rigidly with the $E_F$, whether the independently varying variable was the hole concentration, the Sn concentration, or the temperature. Such behavior is unusual in semiconductors and semimetals where, in the rigid-band model, dopant atom energy levels are defined with respect to the band edges, not to the $E_F$. Therefore, this suggests that the observed doping effect of Sn in Bi is incompatible with the rigid-band model. Here, we show that the mechanism by which Sn generates holes in elemental Bi is actually similar to that described for In and Ga thus far. Although we do not provide the temperature dependence of the band evolution because of the general limitation of the DFT to zero kelvins, it is expected that our doping mechanism could provide a more suitable model for the unexplained temperature dependence of Bi:Sn.[34]

Sn in Bi exerts a slightly stronger negative pressure than In, as far as the local relaxation around the impurity is concerned, but much weaker than Ga (see SOM). The NN and NNN Sn-Bi distances are 3.03 Å and 3.36 Å, respectively. The supercell DOS, including the spin-orbit interaction, is presented in Fig. 5(a) - (c). It shows an HDS peak at -6 eV below $E_F$, more separated from the host Bi states than in Bi:In or Bi:Ga. The local binding behavior, as well as the orbital character of the HDS peak is similar to that in Bi:In, and the absence of a DDS peak near $E_F$ is also seen in Fig. 5(c). Electron counting shows that the HDS accommodates two electrons (one from Sn and the other from six neighboring Bi atoms) per Sn atom. As with Bi:In and Bi:Ga, one Sn atom substituted for one Bi atom forces two valence electrons into the HDS, which creates two holes in the main valence band block. However, since Sn has one more valence electron than Bi, it provides one additional electron to the main valence band block, compensating for one of the two holes created by the formation of the HDS. One substitutional Sn atom ends up contributing one hole to the main valence band block of Bi, and thus it is reasonable to describe Sn as a single hole acceptor in Bi, as demonstrated experimentally[22] at low temperature. The KKR-CPA results, presented in Fig. 5(d) (and Fig. S7 in SOM), again confirm the supercell results. Fig. 5(d) shows a much sharper HDS peak, compared to those seen in the KKR-CPA results for In (Fig. 2(d)) and Ga (see SOM, Fig. S5(d)). Along with the fact that the position of the HDS in Bi:Sn is considerably deeper in energy than in the cases of group III impurities, this leads to less overlap of the HDS with the main valence block in the KKR-CPA results. The DOS near the $E_F$ (Fig. S7 in SOM) confirms the absence of a resonant DDS even for the lowest Sn concentration, as well as the lowering of $E_F$ with increasing Sn concentration, thus indicating that Sn behaves as an acceptor in Bi.

The difference in electron counts between In or Ga doping, and Sn-doping has a second consequence. As discussed earlier, the In or Ga atoms remain electrically neutral in Bi and do not lead to ionized impurity scattering, whereas the same does not hold for the Sn atoms. Sn has four valence electrons, two 5$s$ and two 5$p$ electrons, so there is a difference in electric charge between $Bi^{3+}$ and $Sn^{4+}$, making $Sn^{4+}$ present as ionized impurities in Bi. Indeed, the transport properties of Sn-doped Bi show ionized impurity scattering.[22]

## III. EXPERIMENTAL
### III.1. Indium-doped bismuth single crystals

$Bi_{100-\alpha}In_\alpha$ single crystals with nominal atomic concentrations α = 0, 0.1, 0.5 were grown by a modified horizontal Bridgman technique. High purity zone-refined Bi (initially 99.999%) and In (99.999%) were loaded into quartz ampoules following the stoichiometric ratios, and the ampoules were sealed after being evacuated down to about $10^{-6}$ torr. Then the samples were placed horizontally and melted at 873 K in a box furnace and slowly cooled down in the presence of a temperature gradient, which grew single crystals. X-ray diffraction (XRD) was used to confirm single crystal peaks on oriented disks of the crystals. The samples used for transport measurements were smaller than the ingot. Their actual In concentrations, reported here, were first measured by the x-ray fluorescence (XRF). Some In was expected to segregate in the intermetallic InBi phase during the growth process: the amount of In segregation in each sample was estimated by detecting the InBi peak in the differential scanning calorimetry (DSC) analysis and comparing its heat of melting to that per mole of the InBi phase. Since we could not find any literature values of the heat of melting of the InBi phase, a monophasic InBi sample was made by co-melting followed by quenching and its heat of melting was measured to be $44.09 \times 10^3$ J/kg. The amount of segregated In metal was then calculated from the heat of melting of the InBi phase in the samples and subtracted from the total amount of In from XRF to obtain the final In concentration in each sample. The actual concentrations of In were determined to be α = 0, 0.09, 0.4; we could not determine how much In substituted for Bi, and how much went interstitially in the lattice.

The crystals were cut into two parallelepipeds of approximate dimensions 2.5mm x 1.5mm x 7mm so that heat and charge flux during the transport measurements were oriented in two different crystallographic directions: parallel to the trigonal axis ($z$) and parallel to the trigonal plane ($xy$-plane), respectively. Here, $x$ denotes the binary direction, and $y$ the bisectrix direction. The trigonal axis is perpendicular to the trigonal plane (inset in Fig. 6(a)). SdH oscillations in $\rho_{xx}(B_z)$ were measured using the AC Transport Option in a Physical Properties Measurement System (PPMS) by Quantum Design and a Lakeshore 370 AC bridge. The samples were cooled down to 2 K and $\rho_{xx}(B_z)$ was recorded while sweeping the magnetic field from 0 to 7 T at 50 Oe/s. The Hall coefficients $R_H$ of the doped samples were determined by measuring Hall resistances in $\rho_{xy}(B_z)$ configuration using the same instruments. Hall resistances were measured in both positive and negative magnetic fields. Then, values in one polarity were subtracted from those in the opposite, in order to remove the magnetoresistance component, which is an even function of magnetic field.[48] Thermal conductivity κ, Seebeck coefficient $S$, and $\rho$ were measured simultaneously while slowly sweeping the temperature from 320 K to 2 K using the PPMS'

Thermal Transport Option. We estimate the error on the individual transport properties to be 5%, with the error in $S$ stemming mainly from the size of the thermometry, and in $\rho$ and $\kappa$ from errors in sample geometry. These geometric errors are canceled out in the measurement of $zT$, allowing for an error of 10% in $zT$. At temperatures above 200 K, thermal radiation adversely affects the accuracy of the thermal conductivity measurement, which is overestimated. This leads to an underestimation of $zT$ above 200 K.

### III.1.1. Shubnikov – de Haas and Hall effect measurements on Bi:In

Quantum oscillations (SdH) were clearly identified in $\rho_{xx}(B_z)$ on samples oriented as shown in the inset of Fig. 6(a). Fig. 6(a) shows the traces of SdH oscillations for three single crystalline samples at 2 K. The amplitude of the oscillations decreases as the impurity content increases, indicating a decrease of the Dingle mobility due to increased impurity scattering.[49] In the presence of a strong magnetic field, electrons become less mobile than holes as they have a threefold higher Dingle temperature.[50] Usually, both electron and hole periods can be resolved, but the electron frequency disappears when the field is aligned with the trigonal ($z$) axis because of a resonance between electron and hole frequencies.[6] Since we are mainly interested in hole concentration of the samples in this study, the fact that we resolve only hole oscillations does not cause difficulties. The magnetic field oscillation frequencies $[\Delta(1/B)]^{-1}$ give the cross sectional area of the Fermi surface $A_F$. Assuming that the band structure near $E_F$ is not much affected by the presence of In (the local rigid-band model calculated in Fig. 4), the Fermi energy of holes $E_F$ and the hole concentration $P$ of each sample can be calculated. The results are summarized in Table I.

The Hall resistivity $\rho_{xy}(B_z)$ of the In-doped samples is shown in Fig. 6(b). At low magnetic field, $\rho_{xy}(B_z)$ of both samples shows a negative slope, revealing the presence of high-mobility electrons (inset in Fig. 6(b)). The high-field value of $\rho_{xy}(B_z)$ is dominated by the presence of majority holes. Following Noothoven van Goor,[22] we perform an analysis, given in the SOM, of the Hall coefficient in a two-carrier system, and derive separately the minority electron ($N$) and majority hole ($P$) concentrations. Table II shows the results. The $P$ values acquired from $\rho_{xy}(B_z)$ are in excellent agreement with those from SdH in Table I, which confirms the validity of the local rigid-band model hypothesis, and thus the acceptor nature of In, as well as the validity of the calculations in Fig. 4. Fig. 1 shows $E_F$ of the samples, illustrated on a schematic band diagram of Bi, again assuming that the bands near $E_F$ are unaffected by the In-doping. According to Noothoven van Goor,[22] about $4 \times 10^{18} cm^{-3}$ holes are required to empty the conduction band at the L-point of the Brillouin Zone of Bi. Since the higher doped α = 0.4 sample has only $7.69 \times 10^{17} cm^{-3}$ holes, the observed presence of a small number of electrons is expected.

The doping efficiency of In (i.e. the number of holes per In atom) is estimated to be significantly lower than that of Sn and Pb. Considering that 0.08 at. % Sn introduces $1.5 \times 10^{19}$ cm$^{-3}$ holes in Bi,[47] In-doping appears about 40 times less efficient than Sn doping, and ~2 at. % In (above solubility limit of In in Bi[51]) is needed to empty the conduction band. This indicates that not all In atoms substitute for Bi atoms, likely due to the presence of interstitial impurities in the Bi lattice that compensate the p-type doping action of substitutional In. Such behavior was previously identified for the case of Li-doped Bi.[25] Total energy KKR-CPA calculations were carried out for two types of defects: interstitial In and interstitial Bi. In the first case (i), we assume that some In atoms substitute for Bi, but the rest end up in interstitial positions. In the second case (ii), we assume that all In atoms are in substitutional positions, but push some of Bi atoms into interstitial sites.

Starting with the first case, the total energy of (i$_a$) Bi$_{0.995}$In$_{0.005}$ (all In substitutional) and (i$_b$) Bi$_{0.9965}$In$_{0.0035}$ + In$_{0.0015075}$ (some In interstitial) shows that the energy of case (i$_b$) is only ~2.5 meV/unit cell higher, than case (i$_a$), where all the In are in the substitutional positions. On the other hand, similar comparison for the second case, Bi versus Bi + Bi$_{0.005}$ (interstitial) gives the difference in energy at about 200 meV/unit cell. This shows that the formation of interstitial In defects is quite likely (2.5 meV $\ll k_B T$ at the crystal growth temperature ~544 K), and interstitial Bi are much less expected.

The resulting DOS from the KKR-CPA calculations for those defects are given in the SOM (Fig. S4). Interstitial In gives one electron to the main valence block, and for small In concentrations ~0.1%, $E_F$ moves according to this number. For higher concentrations near the pseudo-gap, interstitial In acts more effectively, and the position of $E_F$ moves approximately as if it had the valence up to three for ~1%. Interstitial Bi is a strong electron donor, with the valence number up to three, and in combination with the substitutional In, this pair behaves as an n-type impurity.

In short, either interstitial In or interstitial Bi leads to the carrier compensation effects and explains the low efficiency observed for In doping in Bi.

III.1.2. Electron scattering mechanism in Bi:In

The temperature dependence of $\rho$ of Bi, $Bi_{99.91}In_{0.09}$, and $Bi_{99.6}In_{0.4}$ along the trigonal axis direction $\rho_{//}(T) = \rho_{zz}(T)$ and in the trigonal plane $\rho_{\perp}(T) = \rho_{xx}(T)$ are shown in Fig. 7(a) and (b), respectively. $\rho_{//}(T)$ and $\rho_{\perp}(T)$ of the In-doped samples show a bump at 45 K, while that of undoped Bi displays a metal-like behavior.[52] This bump is more evident for the higher doped sample. It is noted that $\rho(T)$ of the In-doped samples shows a different behavior from what has been observed in Sn-doped Bi[39] samples doped with almost the same acceptor concentrations. In Sn-doped Bi, $\rho(T)$ flattens out above 200 K at higher Sn concentrations. On the other hand, $\rho(T)$ of the In-doped samples closely follows what has been observed in compensated Bi,[53] where a significant amount of neutral impurity scattering is introduced by doping Bi with almost same concentration of Te and Sn simultaneously. By fitting the experimental data on compensated Bi with the neutral impurity scattering taken into account, Issi *et al.*[53] confirmed that the bumps in $\rho(T)$ at the low temperature is a characteristic of the dominant role of neutral impurity scattering in this system.

Here, we employ a similar calculation procedure for the 0.4% In-doped sample in the trigonal plane (Fig. 7(b)). Issi *et al.*[53] assumed that compensated Bi has the same electron ($N$) and hole ($P$) concentrations as undoped Bi (i.e. $N = P$). In contrast, in In-doped Bi, the carrier concentrations are unequal with $P > N$, and known only at the lowest temperatures where Hall and SdH effects determine them unequivocally (see III.1.1). To interpret the data of Fig. 7(b), we need to determine the temperature dependence of $N(T)$ and $P(T)$. Using Matthiessen's rule for scattering, the total electrical conductivity $\sigma$ (= $1 / \rho$ at zero magnetic field) for each crystallographic direction of the In-doped samples can be expressed as

$$\sigma = \sigma_e + \sigma_h = Nq\left(\frac{1}{\mu_\phi} + \frac{1}{\mu_r}\right)^{-1} + Pq\left(\frac{1}{\nu_\phi} + \frac{1}{\nu_r}\right)^{-1} \qquad (1).$$

where $\sigma_e$ and $\sigma_h$ are the partial electrical conductivities of electrons and holes, $\mu$ and $\nu$ are the mobilities of electrons and holes, respectively, and $q$ is the absolute value of the electron charge. The index $\phi$ refers to the scattering by acoustical phonons, while the index $r$ refers to the scattering due to the added impurity. At 2 K, we may assume that $\sigma$ has reached its residual value $\sigma_r$, hence $1 / \mu_\phi \ll 1 / \mu_r$, and $\sigma \cong \sigma_r = Nq\mu_r + Pq\nu_r$. Since $N$ and $P$ at 2 K are known from above, $\mu_r$ and $\nu_r$ can be obtained by fitting an experimental galvanomagnetic coefficient. Saunders *et al.*[54] showed that several band parameters can be determined simultaneously by measuring only one component of the magnetoresistivity tensor at intermediate magnetic fields, which had been suggested theoretically by Aubrey.[55] The in-plane longitudinal magnetoresistivity $\rho_{xx}(B_z)$ of single-crystal Bi is given as

$$\rho_{xx}(B_z) = \frac{\sigma_{xx}}{\sigma_{xx}^2 + \sigma_{xy}^2}$$

$$= \frac{1}{q} \frac{\left(\dfrac{\bar{\mu}N}{1+\mu_x\mu_y B^2} + \dfrac{\nu P}{1+\nu^2 B^2}\right)}{\left(\dfrac{\bar{\mu}N}{1+\mu_x\mu_y B^2} + \dfrac{\nu P}{1+\nu^2 B^2}\right)^2 + \left(\dfrac{-\mu_x\mu_y NB}{1+\mu_x\mu_y B^2} + \dfrac{\nu^2 PB}{1+\nu^2 B^2}\right)^2} \qquad (2).$$

where $\mu_x$ and $\mu_y$ are the electron mobilities taken for each ellipsoid that constitutes the electron Fermi surface along x- and y-axes, respectively, $\nu = \nu_x = \nu_y$ is the hole mobility in the xy-plane, which is isotropic, and $\bar{\mu} = (\mu_x + \mu_y)/2$. Fig. 7(c) shows the experimental data points of $\rho_{xx}(B_z)$ at 2 K and the fitted curve using Eq. (2) with the known N and P. The method of least squares was used with $\mu_x$, $\mu_y$, and $\nu$ as unknowns for the fitting. Overall, very good agreement between the experimental points and the fitted curve was obtained. Noothoven van Goor[22] also used the same procedure and found that the fitting does not work at relatively higher magnetic fields (say $\mu B > 50$), because, experimentally, the magnetoresistance does not saturate at high magnetic fields. There are several possible reasons why the model does not apply at very high fields: geometrical magnetoresistance[56] and magnetostriction,[57] both of which induce[56,58] an additional magnetoresistance that remains quadratic in field. Other anomalies are observed at very high fields in other transport properties.[15] Since Eq. (2) requires saturation at high fields, we confine the fitting within intermediate magnetic fields. The obtained $\mu_r$ and $\nu_r$ at 2 K are assumed to result from neutral impurity scattering, and thus to be insensitive to temperature, which is the case specifically in Bi.[53] Since we know the temperature dependence of $\mu_\phi$ and $\nu_\phi$,[52,59] the temperature dependence of combined mobilities can be calculated using Matthiessen's rule, as in Eq. (1). Then, N and P at each temperature are found again by fitting $\rho_{xx}(B_z)$ with the known mobilities. Table III summarizes the carrier concentrations as well as the mobilities in the trigonal plane of the 0.4% In-doped Bi sample obtained using the above procedure. It was noticed that the fit to $\rho_{xx}(B_z)$ above 100 K is not as good as at lower temperatures, possibly because the thermal excitations from the valence band at the L-point are not taken into account. Finally, using Eq. (1), the calculated value of $\rho_\perp = 1/\sigma_{xx}$ at each temperature is plotted as a full black line in Fig. 7(b) along with the calculated curve for undoped Bi (dashed black line). Both computed curves fit the experimental data well. This result gives evidence that the additional scattering present in the In-doped Bi samples is indeed temperature independent, as expected from neutral, rather than ionized, impurity scattering. The possible sources of such scattering centers could be grain boundaries formed by InBi segregation, substitutional In atoms that remain electrically neutral, or both.

III.1.3. Thermoelectric transport measurements on Bi:In

In undoped Bi, the Seebeck coefficient (or thermopower) S is always negative in the trigonal axis and trigonal plane directions because the mobility of electrons is much higher than that of holes due to their small effective mass.[11,14] The zT along the trigonal axis direction is larger than that in the trigonal plane[16] in undoped Bi. While the latter is also the case in the In-doped samples, it is interesting that the doped samples show enhanced negative S compared to undoped Bi (Fig. 8(a) and (b)). A model is developed in the SOM that calculates the total thermopower of the samples, based on the fact that there are two carriers, with $S_e$ (< 0) and $S_h$ (> 0), the partial thermopowers of electrons and holes, respectively. Moreover, in a similar way with Matthiessen's rule, the model combines S for different scattering mechanisms (indexed $\phi$ and $r$) using the Gorter-Nordheim rule (see SOM). The resulting total $S$ calculated by Eq. (S4) in SOM is plotted in Fig. 8(b) for both undoped Bi and 0.4% In-doped sample. The computed curves indicate that the increased $S$ arises from an increase in partial $S_e$, itself due to the appearance of neutral impurity scattering, as was the case in compensated Bi[53], but here combined with the decrease in the $E_F$ in the conduction band. The evidence for the effect of neutral impurity scattering is that $S$ in the In-doped Bi samples is larger than that in Sn-doped Bi samples (Ref [39] and III.2 in this work) with a similar carrier concentration at the same temperature. The effect arises because neutral impurity scattering gives a more favorable $\lambda$ than acoustic phonon scattering. The magnitude of enhancement observed experimentally is actually smaller than that shown by the calculation; a discrepancy also observed in the compensated Bi,[53] which is probably related to the unknown temperature-dependence of the band structure of Bi.[60]

Figure 9(a) and (b) show the thermal conductivity $\kappa(T)$ of Bi, $Bi_{99.91}In_{0.09}$ and $Bi_{99.6}In_{0.4}$ in both trigonal axis direction and trigonal plane, respectively. Interestingly, there is almost no difference in $\kappa$ between undoped Bi and the In-doped samples over the measured temperature range, even though a certain degree of decrease was expected in the doped samples due to the impurity scattering. It is possible that noticeable difference could exist around the dielectric maximum below 10 K where the phonon mean free path is most sensitive to the impurity scattering. In our measurements, the samples conduct heat so well in that temperature range, that the experimental error bars become too large to draw any conclusions below 10 K. The $zT^4$ of the samples in the trigonal axis ($zT_{//}$) and trigonal plane ($zT_\perp$) directions is shown in Fig. 9(c) and (d), respectively. In both directions, the In-doped samples show higher $zT$ than undoped Bi, mainly due to the enhanced $S$. While $zT_\perp$ is still low, the $zT_{//}$ of the In-doped samples reaches 0.36 at 280 K, which is a 20% enhancement compared to that of undoped Bi.

The thermopower in a strong magnetic field ($\mu B >> 1$) becomes independent of the relaxation time, and thus of the type of scattering.[24] $S(T)$ measured in the 7 T magnetic field is shown in Fig. 10. The effects of the neutral impurity scattering marked in the zero magnetic field disappear in the 7 T field, which makes it possible in principle to isolate the effect of scattering on the zero-field thermopower. However, this separation is complicated by the fact that electron and hole mobilities are different and therefore the magnetoconductivities in Eq. (S2) in the SOM also have a field dependence. In the zero magnetic field, $S(T)$ is always negative in both measurement directions (Fig. 8) as a result of higher mobility of electrons than that of holes. When a magnetic field is applied, the magnetoconductivity of the electrons decreases rapidly, thus the partial thermopower of the holes is better revealed. Especially, since the hole mobility is much higher in the trigonal plane than in the trigonal axis direction,[59] the hole partial thermopower contribution is more in evidence in $S_\perp$ than in $S_{//}$ (Eq. (S2) in SOM).

Fig. 10(a) and (b) show that this qualitative prediction explains the experimental data. In the trigonal axis direction (Fig. 10(a)), the electron mobility is still dominant, leading to the mostly negative $S_{//}$ in all three samples except below 10 K, where their $S_{//}$ changes sign to positive. The 0.4% In-doped sample maintains positive $S_{//}$ up to 30 K. Note that all samples exhibit positive phonon-drag thermopower peaks at the lowest temperatures, indicating the prevailing hole transport in this temperature regime; this is true also of the undoped samples, as in Issi's review.[14] More pronounced p-type behaviors are observed in the trigonal plane (Fig. 10(b)). Because of the higher mobility of holes in this direction, $S_\perp$ of both In-doped samples remains positive over the whole measured temperature range. The 0.4% In-doped sample has a maximum at 140 K then its $S_\perp$ decreases as temperature rises because of the thermal excitation of minority carriers. The same behavior is observed in the 0.09% In doped sample, except that this sample has an additional maximum at $T < 50$ K. It is possible that in the lower doped sample, $E_F$ hits two different valence bands with different band masses as temperature changes, while the higher-doped sample has $E_F$ deeper in the valence bands where it is less affected by the relative changes in band energies due to temperature. $S_\perp$ of undoped Bi remains negative at most of temperatures. However, the positive $S_\perp$ at the low temperature is more pronounced compared to that in the other direction, again because of the higher hole mobility in this direction. These observations of $S(T,H=7Tesla)$ confirm that In enhances the p-type transport in Bi, consistent with the results of the SdH and Hall measurements.

### III.2. Tin-doped bismuth single crystals

Many experimental studies on the Bi:Sn system exist.[22,34,39,46,47] We add data on a single crystal of Bi:Sn with ~5.5 X $10^{17}$ cm$^{-3}$ holes to provide a direct comparison with the similarly-doped Bi:In samples prepared in an identical way. DSC measurements revealed no Sn segregation in the sample down to 5 ppm. The hole concentration was determined by SdH and Hall effect measurements at 2 K. Its $\rho$ and $S$ are reported in Fig. 11 both parallel and perpendicular to the trigonal axis; the data for the two Bi:In samples (3.9 x $10^{17}$ cm$^{-3}$ and 7.7 x $10^{17}$ cm$^{-3}$), are repeated for comparison. While the hole concentration of the Bi:Sn sample is in-between those of the Bi:In samples, its $\rho$ in both directions is close to that of the lower doped Bi:In sample below 100 K and is between those of the two Bi:In samples at higher temperatures (Fig. 11(a) and (b)). The higher doped Bi:In sample shows the highest $\rho$ over the measured temperature range. The mobility of Bi:In is thus lower than Bi:Sn, and rapidly decreases as In concentration increases. This is presumably due to defect scattering, which can be due to the compensation effects identified to exist in Bi:In. Regarding $S$, we do not see any enhancement of $S$ in Bi:Sn compared to undoped Bi, which we have seen in Bi:In (Fig. 8). The thermal conductivity $\kappa$ of the Bi:Sn sample (not reported) was measured to be indistinguishable from that of the Bi:In samples. Fig. 12 (a) and (b) show that the figure of merit $zT$ of the Bi:In samples is improved over that of the similarly doped Bi:Sn sample in both directions. At 300 K, the improvement is by ~ 40 % in $zT_{//}$, and by ~ 21 % in $zT_\perp$, respectively. The increase in $zT$ is mainly due to the enhanced $S$, which is caused by the neutral impurity scattering. Note that neutral impurity scattering also leads to an increase in $zT$ in Bi doped with compensating Te and Sn, which also act as neutral scatterers,[53] while Goldsmid[61] established that no such result can be obtained from ionized impurity scattering. This result suggests that indeed the isovalent doping mechanism can also be utilized for improving the thermoelectric efficiency of the host material.

### IV. CONCLUSIONS

In summary, this study describes a new concept of isovalent doping, using as paradigm the behavior of trivalent group III elements in the elemental semimetal Bi, in which the Bi atoms are also mainly trivalent. The acceptor behavior of In and Ga in Bi are identified using supercell calculations and confirmed using KKR-CPA electronic structure calculations. The doping mechanism arises from the formation of a hyperdeep defect state about 5 eV below the Fermi energy, where the *5s* electrons of In hybridize with the *6p* electrons of the neighboring Bi atoms, effectively localizing these Bi electrons and depleting the electron count of the rest of the solid near the Fermi energy. This method also leaves the group III element charge-neutral, which, in principle, avoids ionized impurity scattering of the free carriers around the Fermi energy. Tetravalent Sn in Bi is not a simple monovalent acceptor, as it was previously believed, and its behavior can be explained by the new doping mechanism proposed here. Experimentally, SdH and Hall measurements are consistent, and confirm the acceptor behavior of In in Bi, as does the measurement of the Seebeck coefficient in a strong magnetic field. Data on Bi:Ga (in SOM) confirm the calculation. In practice, In behaves experimentally like a partially compensated acceptor with a much lower doping efficiency than the calculation predicts. Interstitial In defects are calculated to act as donors, and are a likely source for this compensation. The thermoelectric and galvanomagnetic measurements prove that In-doping leads to temperature (and energy)-independent neutral impurity scattering in Bi, which results in an enhancement of the Seebeck coefficient over those of undoped Bi and Sn-doped Bi with a similar hole concentration. The thermoelectric figure of merit of the In-doped Bi samples shows 20% and 40% improvements over those of undoped Bi and Sn-doped Bi, respectively. Therefore, the isovalent doping we introduced here can lead to better thermoelectric materials. The fact that, in principle, isovalent doping avoids ionized impurity scattering may well lead to the search for similar impurities in technologically relevant semiconductors, where they may minimize mobility loss due to doping.

## ACKNOWLEDGMENT

Support for this work comes from AFOSR MURI FA9550-10-1-0533, "Cryogenic Peltier Cooling". B. W. was further partly supported by the National Science Center (Poland) (Project No. DEC-2011/02/A/ST3/00124), and by an allocation of computing time from the Ohio Supercomputer Center.

**TABLES**

| Sample | $[\Delta(1/B)]^{-1}$ (T) | $A_F$ ($10^{16}$m$^{-2}$) | $E_F$ (meV) | $P$ ($10^{17}$cm$^{-3}$) |
|---|---|---|---|---|
| $\alpha = 0$ | 6.352 | 6.06 | 11.5 | 2.93 |
| $\alpha = 0.09$ | 7.665 | 7.31 | 13.9 | 3.88 |
| $\alpha = 0.4$ | 12.09 | 11.5 | 21.9 | 7.69 |

TABLE I. Magnetic field oscillation frequencies $[\Delta(1/B)]^{-1}$, cross sectional areas of Fermi surface $A_F$, Fermi energy $E_F$, and hole concentrations $P$ of the Bi$_{100-\alpha}$In$_\alpha$ samples obtained from the SdH oscillations at 2 K.

| Sample | $N$ ($10^{17}$cm$^{-3}$) | $P - N$ ($10^{17}$cm$^{-3}$) | $P$ ($10^{17}$cm$^{-3}$) |
|---|---|---|---|
| $\alpha = 0.09$ | 0.74 | 3.2 | 3.94 |
| $\alpha = 0.4$ | 0.58 | 6.5 | 7.08 |

TABLE II. Electron ($N$), excess hole ($P$-$N$), and hole ($P$) concentrations calculated from the $\rho_{xy}(B_z)$ measurement at 2 K.

| $T$ (K) | $N$ ($10^{17}$cm$^{-3}$) | $P$ ($10^{17}$cm$^{-3}$) | Mobility ($10^4$ cm$^2$V$^{-1}$s$^{-1}$) | | |
|---|---|---|---|---|---|
| | | | $\mu_x$ | $\mu_y$ | $v$ |
| 2 | 0.58 | 7.69 | 184 | 1.66 | 1.4 |
| 10 | 0.61 | 7.7 | 171 | 1.58 | 1.4 |
| 20 | 0.68 | 7.74 | 146 | 1.28 | 1.38 |
| 40 | 0.96 | 8.08 | 99 | 0.91 | 1.35 |
| 60 | 1.41 | 8.33 | 68.8 | 0.62 | 1.31 |
| 80 | 2.18 | 8.53 | 47.5 | 0.44 | 1.23 |
| 100 | 3.22 | 9.09 | 33.5 | 0.33 | 1.16 |
| 140 | 5.74 | 13.83 | 18 | 0.19 | 1 |
| 180 | 9.58 | 16.57 | 10.4 | 0.15 | 0.82 |
| 220 | 13.55 | 18.41 | 6.84 | 0.09 | 0.65 |
| 260 | 18.32 | 20.24 | 4.49 | 0.06 | 0.57 |
| 300 | 23.3 | 28.6 | 3.15 | 0.04 | 0.42 |

TABLE III. Temperature dependence of parameters in the trigonal plane of Bi$_{99.6}$In$_{0.4}$ sample obtained by fitting Eq. (2) to the experimental data: $N$, electron concentration; $P$, hole concentration; $\mu_x$, binary electron mobility; $\mu_y$, bisectrix electron mobility; $v$, isotropic hole mobility in the trigonal plane.

**FIGURE CAPTIONS**

FIG. 1. (Color online) Schematic band diagram indicating positions of the electronic bands of Bi and the Fermi energy $E_F$ for pure Bi and for the Bi:In samples further in this study.

FIG. 2. (Color online) Density of states (DOS) for $Bi_{0.99}In_{0.01}$. (a-c) FP-LAPW supercell results, (d) KKR-CPA result. Total DOS in (a-c) is divided by the number of atoms in the supercell (96) to allow for direct comparison with (d). The DOS of the impurity In atom is not multiplied by the concentration.

FIG. 3. (Color online) Color map of the charge density (e/Å$^3$) in logarithmic scale around an In impurity, corresponding to the hyperdeep defect state (HDS) peak in Fig. 3(a-b). The nearest (Bi1) and next nearest (Bi2) atoms are labeled.

FIG. 4. (Color online) Evolution of density of states (DOS) for the Bi:In from the KKR-CPA calculations. The In and Bi atomic contributions are given per atom, not multiplied by the concentrations. P-type behavior of In in Bi is clearly shown. Modifications of DOS are close to, but not exactly rigid-band like.

FIG. 5. (Color online) Density of states (DOS) for $Bi_{0.99}Sn_{0.01}$. (a-c) FP-LAPW supercell results, (d) KKR-CPA result.

FIG. 6. (Color online) (a) Raw Shubnikov – de Haas (SdH) traces for $Bi_{100-\alpha}In_\alpha$ single crystalline samples ($\alpha$ = 0, 0.09 and 0.4) versus inverse field (1/B), at 2 K and with the background subtracted. The inset shows the configuration used for the SdH measurements which we denote $\rho_{xx}(B_z)$ where the first, second, and third indexes indicate the direction of current flow, potential difference, and magnetic field, respectively. Each axis in the configuration corresponds to binary ($x$), bisectrix ($y$), and trigonal ($z$) crystallographic direction, respectively. (b) Hall resistivity $\rho_{xy}(B_z)$ versus magnetic field for $Bi_{100-\alpha}In_\alpha$ samples measured at 2 K. The symbols are: (green cross) α=0.09, and (blue circle) α=0.4. The points indicate the experimental data, while the lines are added to guide the eye. The inset contains magnification of $\rho_{xy}(B_z)$ at low magnetic field, which shows the transition from the negative slope to the positive slope for both samples.

FIG. 7. (Color online) Temperature dependence of the electrical resistivity of undoped Bi and Bi:In single crystalline samples in (a) the trigonal axis direction ($\rho_{//}$), and (b) the trigonal plane ($\rho_\perp$). The symbols are the experimental data: (red diamond) undoped Bi, (green cross) $Bi_{99.91}In_{0.09}$, and (blue circle) $Bi_{99.6}In_{0.4}$. The solid black line represents the computed curve for $Bi_{99.6}In_{0.4}$, while the dashed black line corresponds to the computed curve for undoped Bi. (c) Magnetoresistivity $\rho_{xx}(B_z)$ of $Bi_{99.6}In_{0.4}$ versus magnetic field measured at 2 K. The points indicate the experimental data, while the solid black curve is the fit generated by the method of least squares with regard to Eq. (2).

FIG. 8. (Color online) Temperature dependence of the Seebeck coefficient of undoped Bi and Bi:In single crystalline samples in (a) the trigonal axis direction ($S_{//}$), and (b) the trigonal plane ($S_\perp$). The symbols are the experimental data: (red diamond) undoped Bi, (green cross) $Bi_{99.91}In_{0.09}$, and (blue circle) $Bi_{99.6}In_{0.4}$. The solid black line represents the computed curve for $Bi_{99.6}In_{0.4}$; the dashed black line corresponds to the computed curve for undoped Bi.

FIG. 9. (Color online) Temperature dependence of the thermal conductivity of undoped Bi and Bi:In single crystalline samples in (a) the trigonal axis direction ($\kappa_{//}$), and (b) the trigonal plane ($\kappa_\perp$). Figure of merit in (c) the trigonal axis direction ($zT_{//}$), and (d) the trigonal plane ($zT_\perp$). The symbols are the experimental data: (red diamond) undoped Bi, (green cross) $Bi_{99.91}In_{0.09}$, and (blue circle) $Bi_{99.6}In_{0.4}$.

FIG. 10. (Color online) Temperature dependence of the Seebeck coefficient of undoped Bi and Bi:In single crystalline samples in 7T magnetic field in (a) the trigonal axis direction ($S_{//}$), and (b) the trigonal plane ($S_\perp$). The direction of magnetic field is parallel to that of heat flux. The symbols are the experimental data: (red diamond) undoped Bi, (green cross) $Bi_{99.91}In_{0.09}$, and (blue circle) $Bi_{99.6}In_{0.4}$. The dashed lines indicating zero Seebeck coefficient are added to help readers recognize the sign change of the Seebeck coefficient.

FIG. 11. (Color online) Temperature dependence of the electrical resistivity in (a) the trigonal axis direction ($\rho_{//}$), and (b) the trigonal plane ($\rho_{\perp}$), and temperature dependence of the Seebeck coefficient in (c) the trigonal axis direction ($S_{//}$), and (d) the trigonal plane ($S_{\perp}$) for Bi:Sn and Bi:In single crystalline samples. The symbols are the experimental data: (orange triangle) Bi:Sn with $5.5 \times 10^{17}$ $cm^{-3}$ holes, (green cross) $Bi_{99.91}In_{0.09}$ with $3.9 \times 10^{17}$ $cm^{-3}$ holes, and (blue circle) $Bi_{99.6}In_{0.4}$ with $7.7 \times 10^{17}$ $cm^{-3}$ holes.

FIG. 12. (Color online) Temperature dependence of the figure of merit in (a) the trigonal axis direction ($zT_{//}$), and (b) the trigonal plane ($zT_{\perp}$) for Bi:Sn and Bi:In single crystalline samples. The symbols are the experimental data: (orange triangle) Bi:Sn with $5.5 \times 10^{17}$ $cm^{-3}$ holes, (green cross) $Bi_{99.91}In_{0.09}$ with $3.9 \times 10^{17}$ $cm^{-3}$ holes, and (blue circle) $Bi_{99.6}In_{0.4}$ with $7.7 \times 10^{17}$ $cm^{-3}$ holes.

**Figure 1 (one column)**

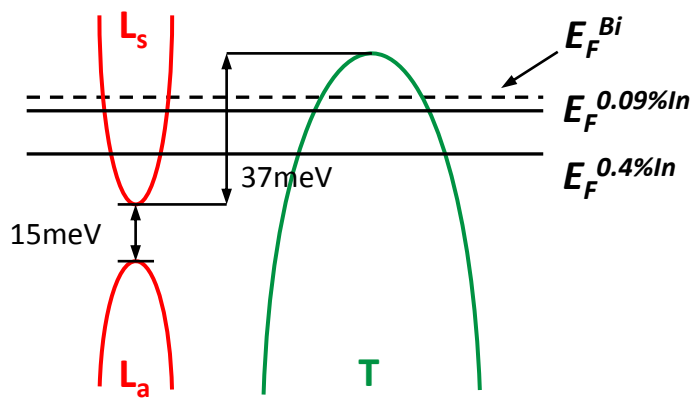

**Figure 2 (two columns)**

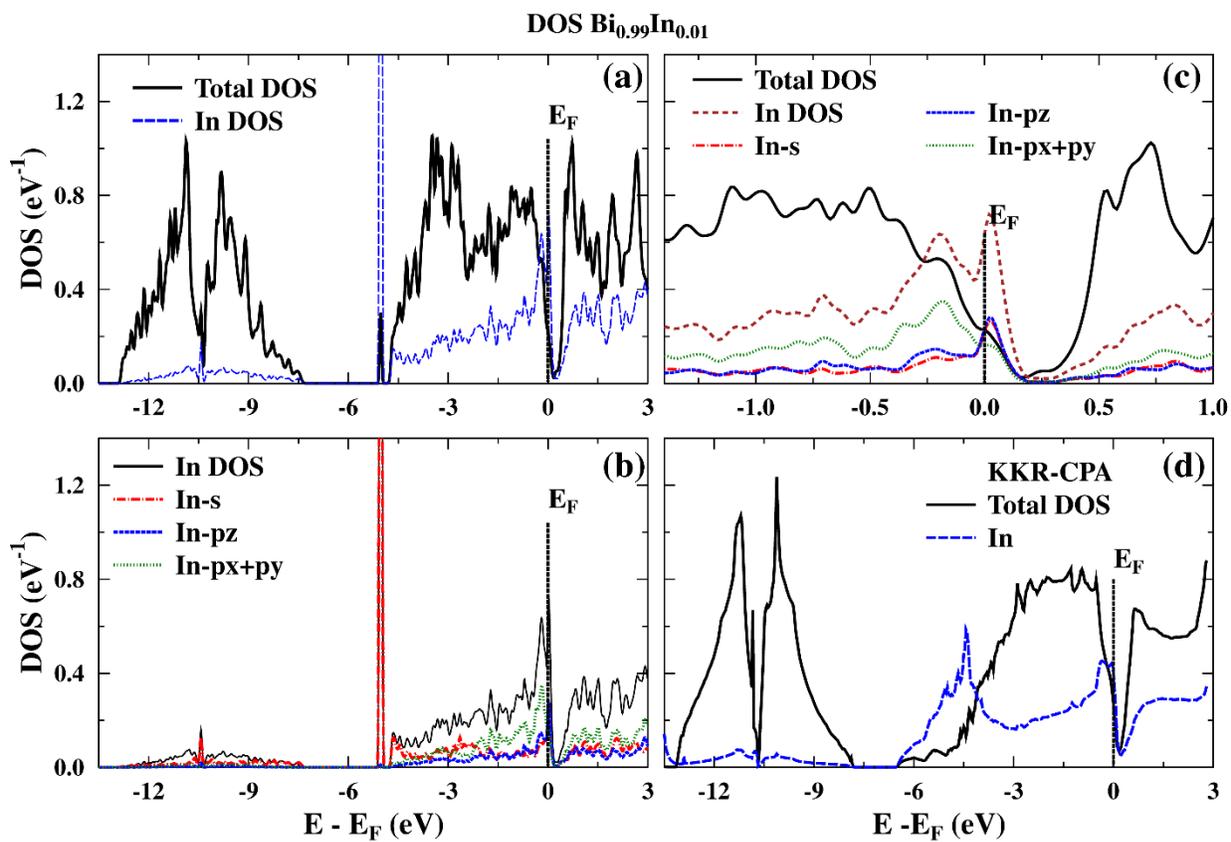

**Figure 3 (one column)**

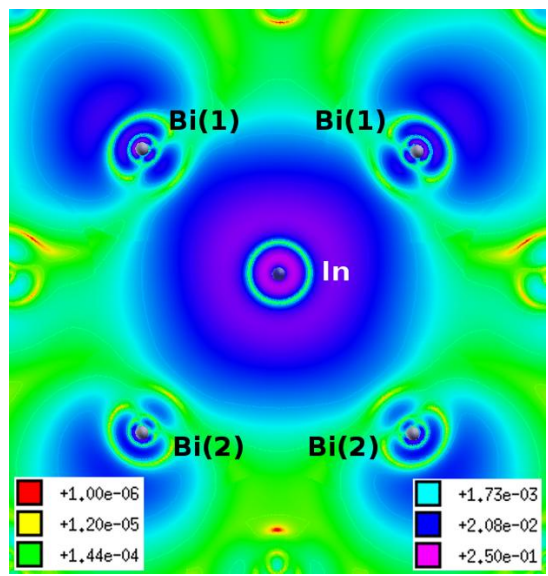

**Figure 4 (two columns)**

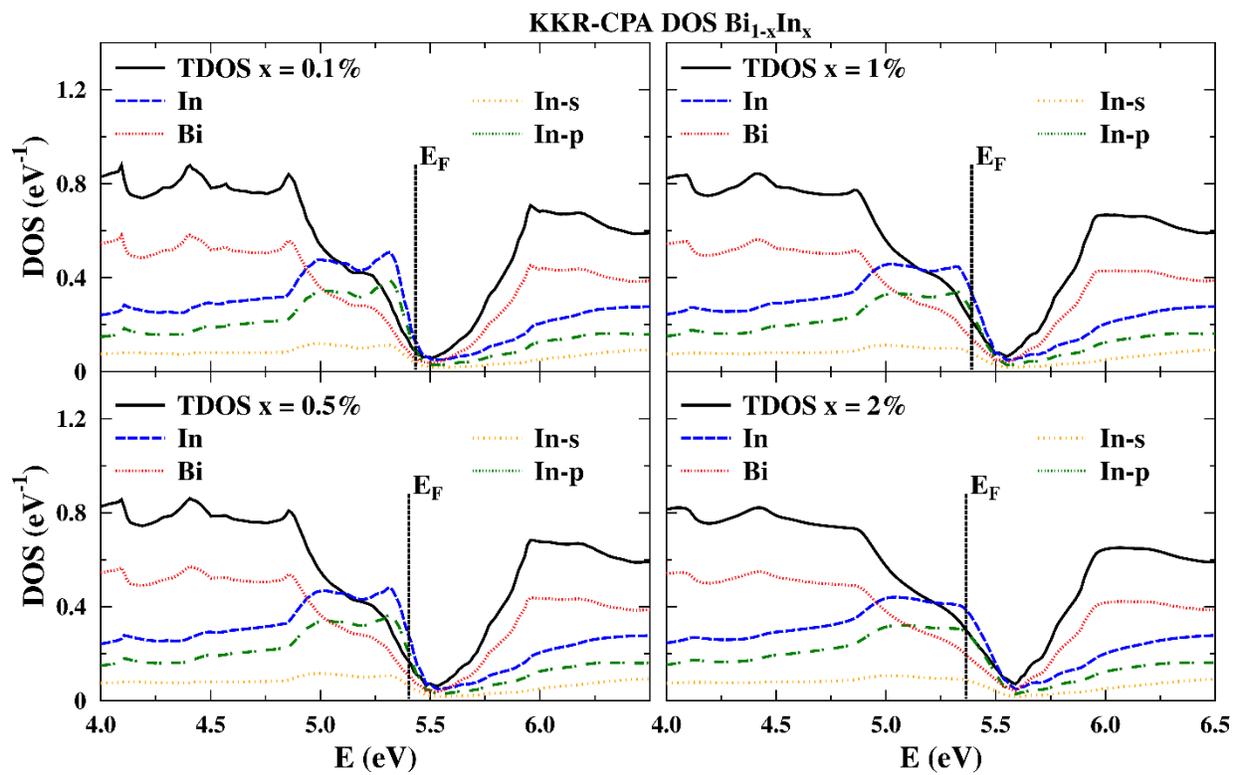

**Figure 5 (two columns)**

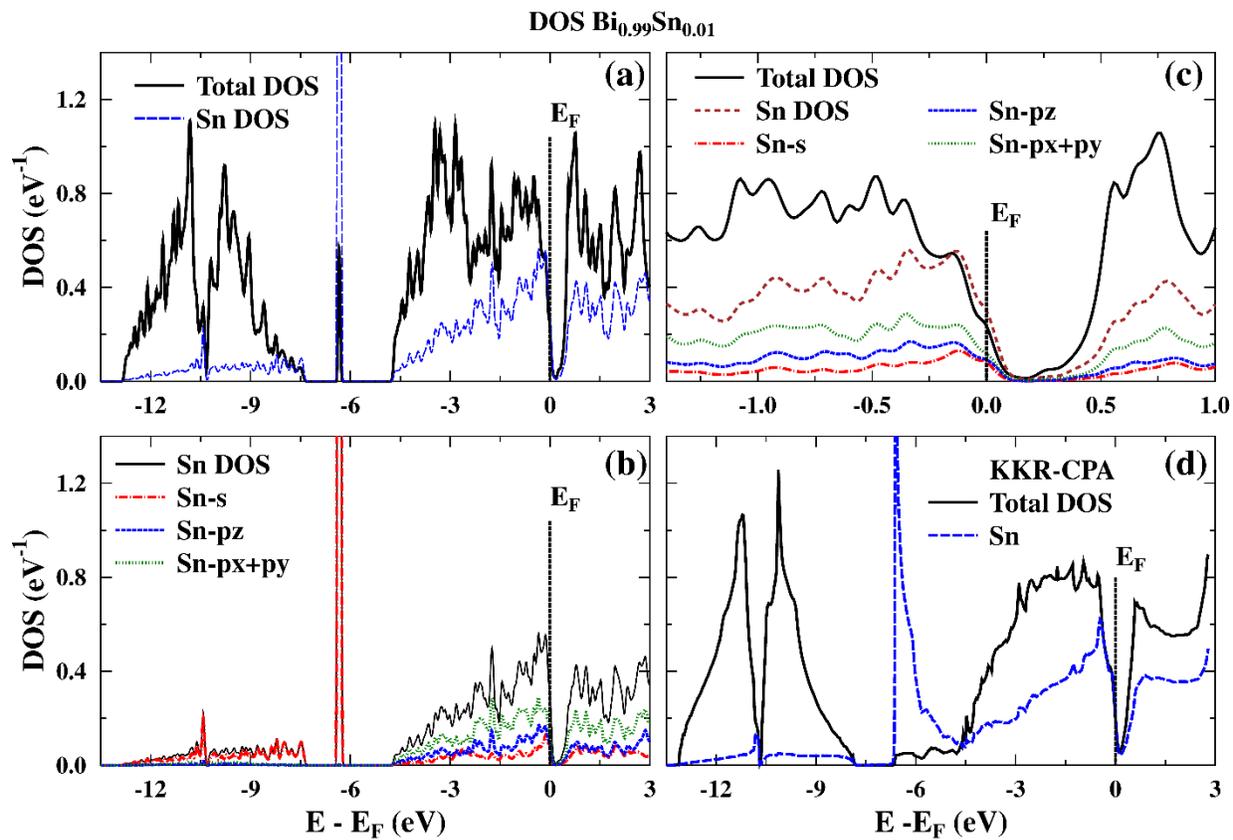

**Figure 6 (one column)**

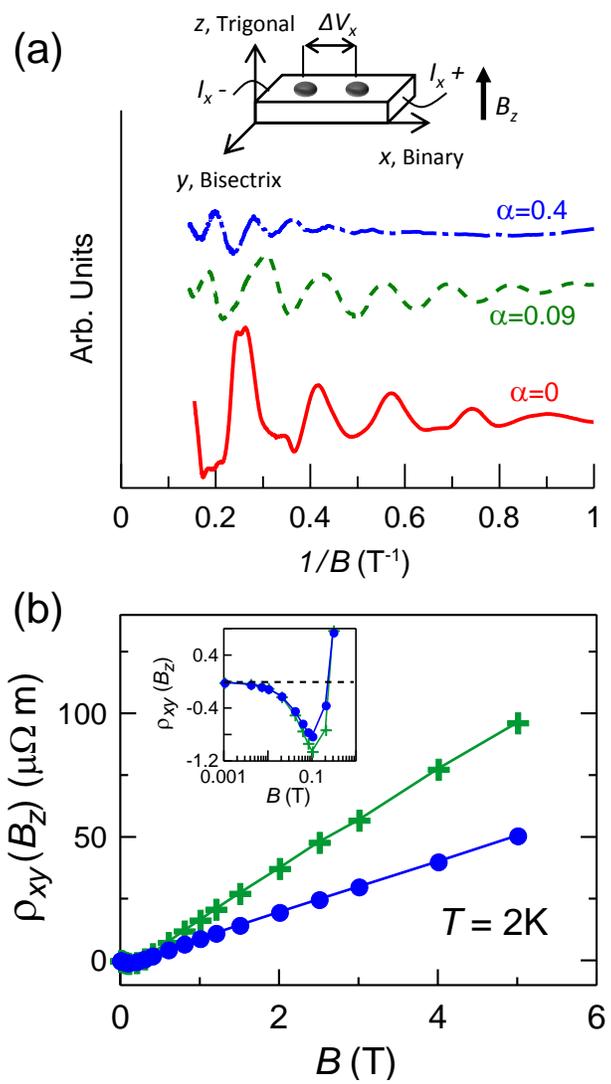

**Figure 7 (one column)**

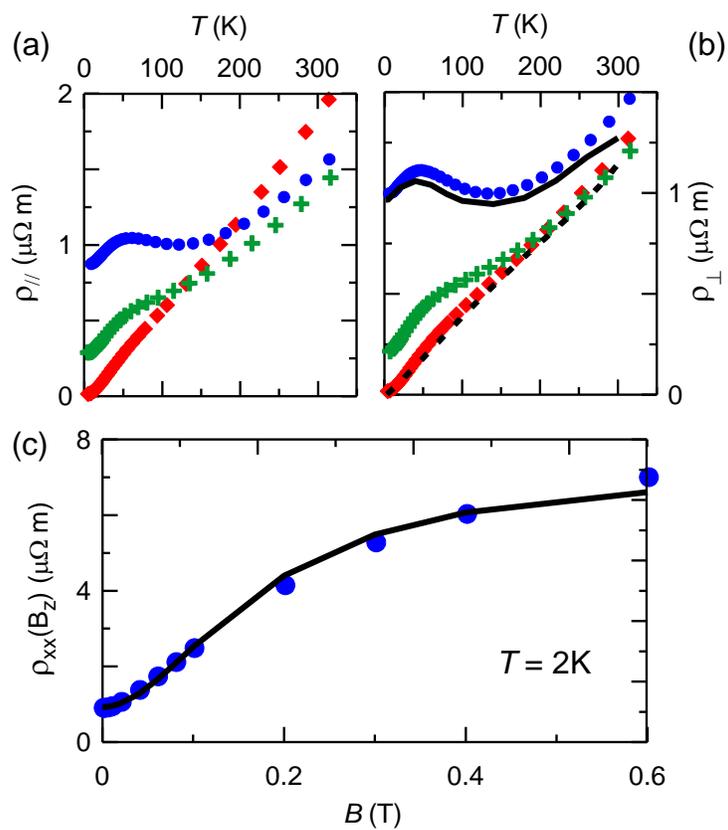

**Figure 8 (one column)**

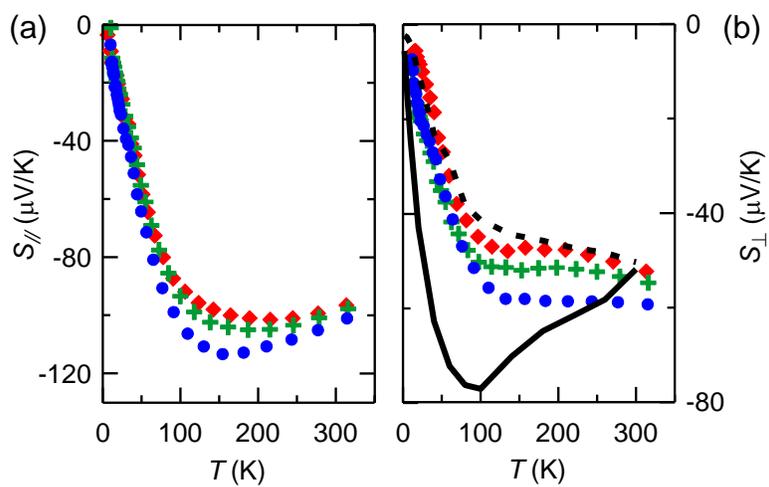

**Figure 9 (one column)**

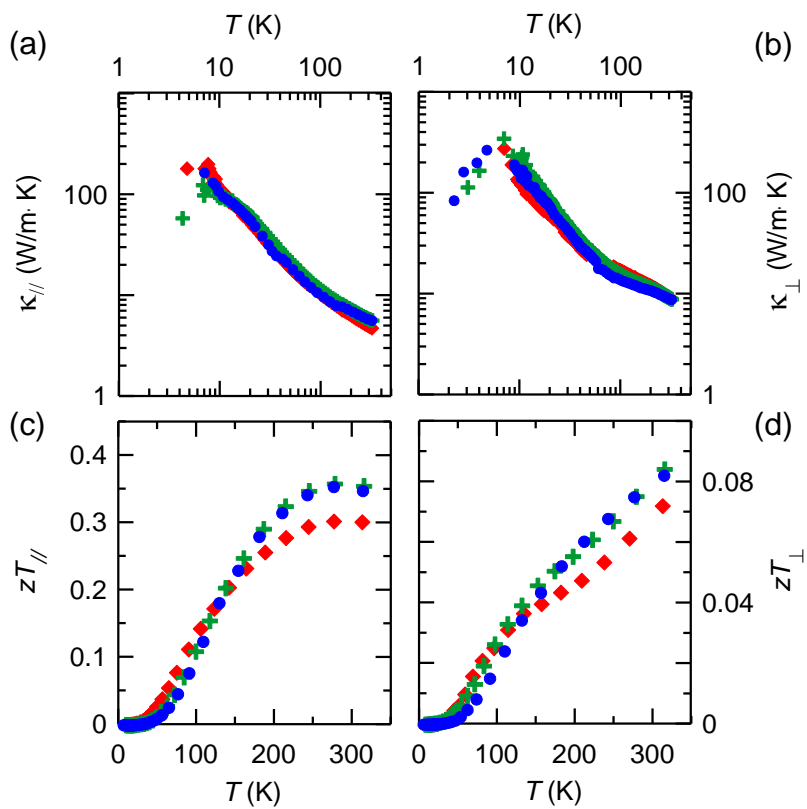

**Figure 10 (one column)**

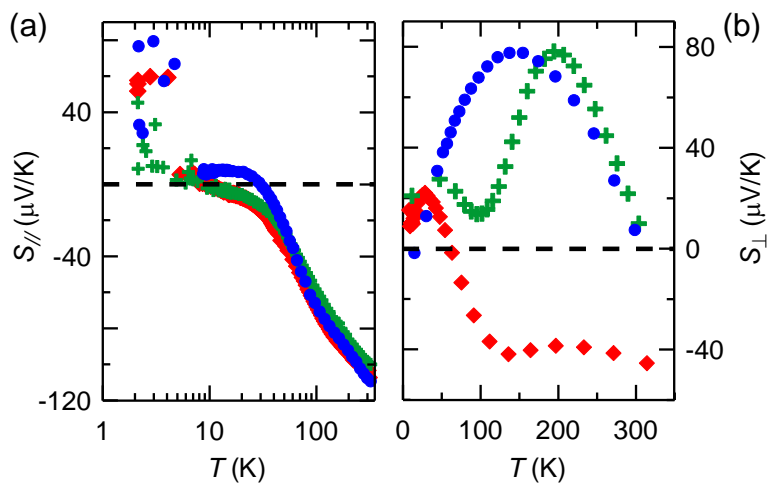

**Figure 11 (one column)**

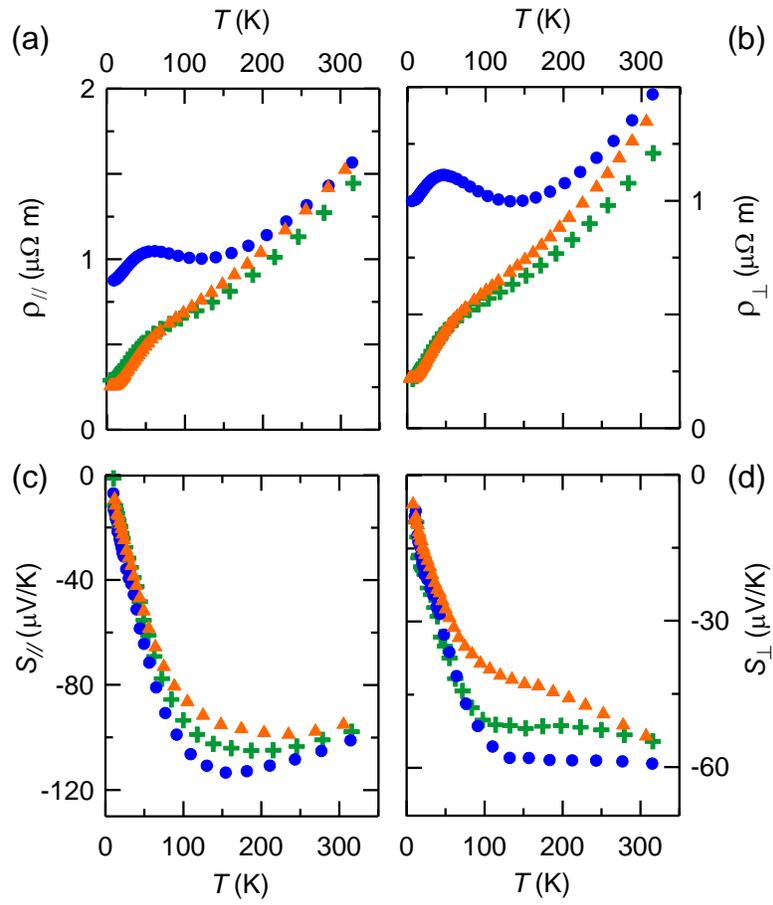

**Figure 12 (one column)**

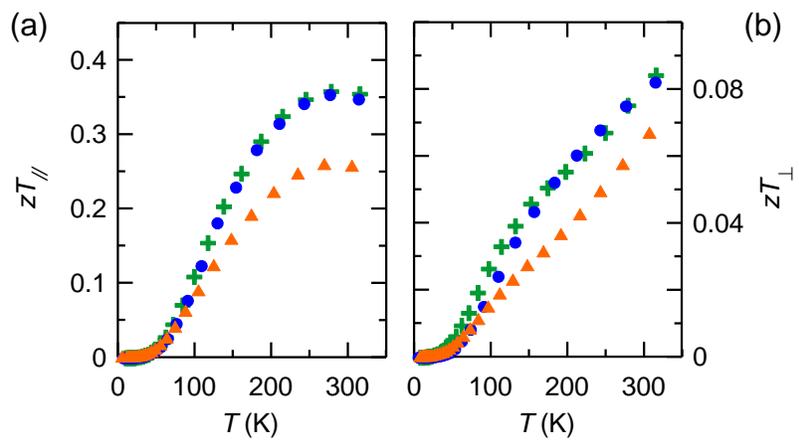

**Supplemental Online Material to:**

**P-type doping of elemental bismuth with indium, gallium and tin: a novel doping mechanism in solids**

Hyungyu Jin[1], Bartlomiej Wiendlocha[1,2], and Joseph P. Heremans[1,3,4]

1. Department of Mechanical and Aerospace Engineering, The Ohio State University, Columbus, OH, USA
2. Faculty of Physics and Applied Computer Science, AGH University of Science and Technology, Al. Mickiewicza 30, 30-059 Krakow, Poland
3. Department of Physics, The Ohio State University, Columbus, OH, USA
4. Department of Materials Science and Engineering, The Ohio State University, Columbus, OH, USA

**Table of Contents**

# 1. Theory

## 1.1. Details of the calculations

Experimental low temperature Bi crystal structure and lattice parameters for the A7 rhombohedral primitive cell[1] were used in the computations: space group no. 166, *R-3m,* a = 4.533Å, c = 11.797Å (in hexagonal notation), Bi atoms positions: (0.234, 0.234, 0.234) and (0.766, 0.766, 0.766). For the supercell full potential linearized augmented plane wave (FP-LAPW) calculations for In, Ga, and Sn doped Bi, the supercell was constructed starting from the hexagonal equivalent of the Bi unit cell (containing 6 Bi atoms). This cell was multiplied, building a 4x4x1 supercell (containing 96 atoms, dimensions: a = 18.13Å, c = 11.80Å), whose size is equivalent to 48 primitive A7 cells. Substitution of a single Bi atom with an impurity atom results in the impurity concentration of 1.04% ≈ 1%, and causes lowering of the unit cell symmetry to *P3m1* (space group no. 156), reducing the number of symmetry operations from 12 to 6. The Perdew-Burke-Ernzerhof Generalized Gradient Approximation[2] (PBE-GGA) was used to calculate exchange-correlation potential. In the first step, atomic positions were relaxed in a semi-relativistic computation on a 3x3x4 **k**-point mesh, and it was observed that all of the impurities (In, Ga, Sn) create a negative chemical pressure, as described in the paper. Final calculations, which included a spin-orbit coupling term, were done on a 3x3x5 **k**-point mesh.

To independently verify the results of the supercell calculations, and to study the effect of the impurity concentration on the evolution of density of states (DOS) near the Fermi energy ($E_F$), the Korringa-Kohn-Rostoker method, with the coherent potential approximation (KKR-CPA) was used. The advantage of the KKR-CPA is that any impurity concentration may be calculated using the same computational geometry, i.e. a primitive rhombohedral Bi crystal cell, however neglecting the local crystal relaxation effects. The KKR-CPA computations were limited to the spherical potential and semi-relativistic approximations. The DOS calculated with KKR-CPA shows good agreement with that obtained from the full potential relativistic supercell calculations. In all KKR calculations, the local density approximation (LDA) was used. The position of $E_F$ in doped samples was obtained using the generalized Lloyd formula,[3] and to increase the unit cell filling, two empty spheres were added between Bi atoms in positions along the trigonal axis: 1a (0,0,0) and 1b (0.5,0.5,0.5). Also using the KKR-CPA, we explored the behavior of possible defects in the Bi:In system, especially the role of interstitial defects.

For all calculations, high convergence limits were used on the self-consistent cycle ($10^{-4}$ q for charge, $10^{-4}$ Ry for $E_F$ and $10^{-6}$ Ry for the total energy). All results were checked for convergence against angular momentum cutoff and **k**-point number.

## 1.2. Elemental bismuth: theory

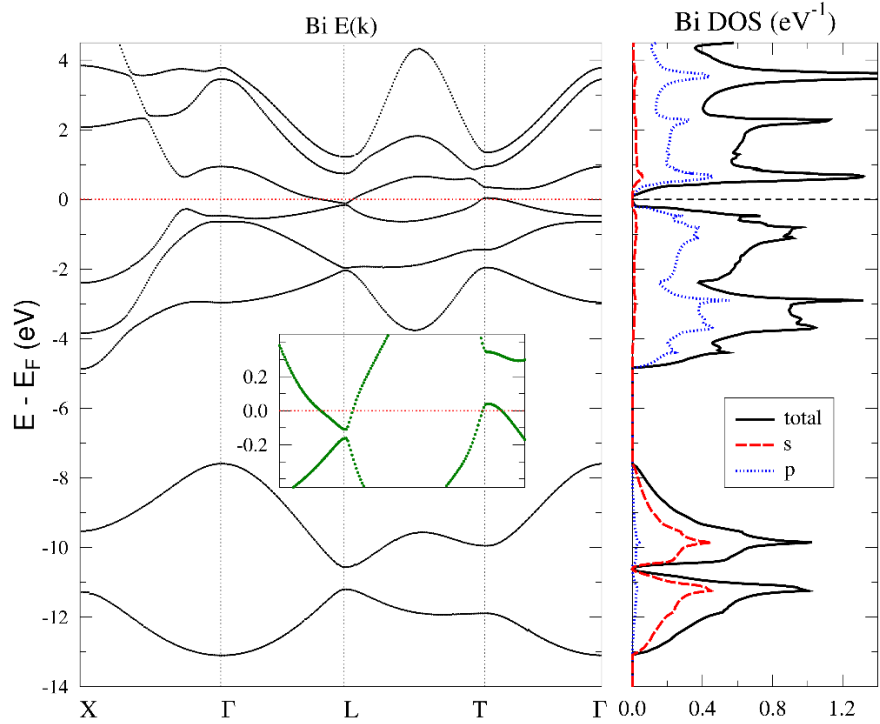

*Figure S1. Electronic band structure of rhombohedral Bi from relativistic FP-KKR calculations. Inset shows the electron and hole bands near the Fermi energy $E_F$.*

For undoped Bi, both FP-LAPW with spin-orbit coupling and fully relativistic (i.e., based on the Dirac equation) full potential KKR[4] (FP-KKR) methods were used, giving very similar results. Electronic dispersion curves and DOS of Bi calculated by the relativistic FP-KKR are presented in Fig. S1. The two low energy-lying bands are occupied mainly by 6s electrons (two pers each band, since there are two Bi atoms per primitive A7 cell), and the main valence block, consisting of three 6p-like bands, accommodates three 6p electrons per Bi atom. The conduction bands are separated from the valence block by the pseudo-gap. Our results are in overall good agreement with experimental finding[5] and the previously published work.[6,7] Relativistic effects in the band structure are visible as, for example, the splitting of the two highest valence bands at the Γ-point, while DOS near $E_F$ is not affected much by the spin-orbit coupling. Our calculations were not able to avoid the common LDA band-gap problem, which here manifests itself as an error in the value of the band overlap. Our calculated value of the L-T overlap is about 170 meV as the bottom of the electron band at the L-point is too low and the top of the hole band at the T-point is too high in energy. A similar value (163 meV) was obtained in the previous linear muffin-tin orbital (LMTO) calculations[7], while the pseudopotential calculations[6] reproduced the experimental 40 meV overlap value; the authors of that work noted this was unexpected. Considering that the common DFT+LDA error for band gaps can be of the order of 0.5 eV (e.g. in Si[8]), and that the absolute error in the band overlap here is smaller (0.13 eV), we find these values satisfactory enough to qualitatively discuss the behavior of impurities in doped systems. More rigorous quantitative analysis would require going beyond the local density methods.

1.3. Indium-doped bismuth: theory

Figure S2 shows the DOS of the nearest (NN) and next-nearest (NNN) neighbor Bi atoms to an In impurity atom. The appearance of the sharp DOS peaks around -5 eV proves that Bi electrons also contribute to the formation of the hyperdeep defect state (HDS). Since the HDS accommodates one electron from In and 1/6 of electron per each of the 3 NN and 3 NNN Bi atoms (thus one Bi electron in total) as described in the main paper (II.1), we can state that both In and Bi equally contribute to the HDS. Figure S3 shows an In impurity atom, 3 NN and 3 NNN Bi atoms, and the plane on which the charge density is projected in Fig. 3 of the main paper.

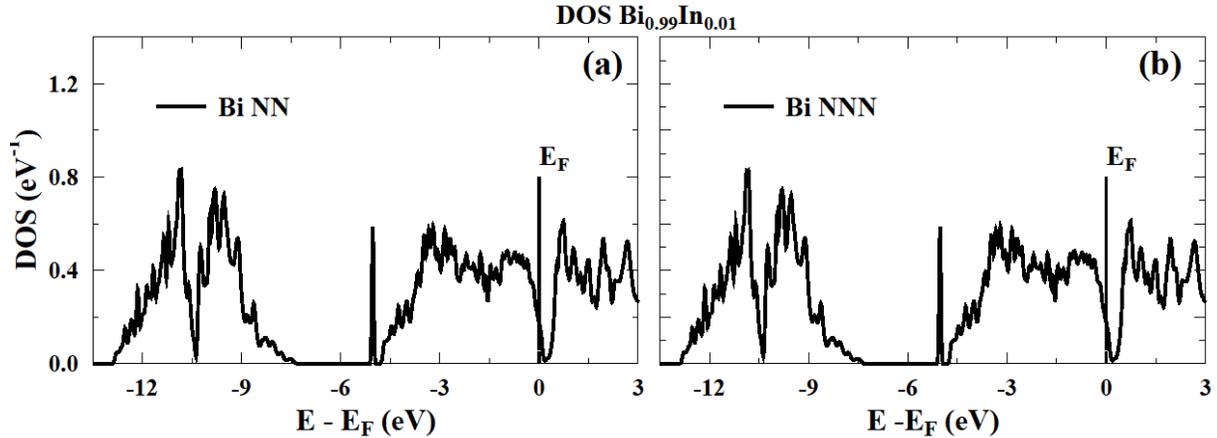

*Figure S2. DOS of the (a) nearest and (b) next-nearest neighbor Bi atoms to an In impurity.*

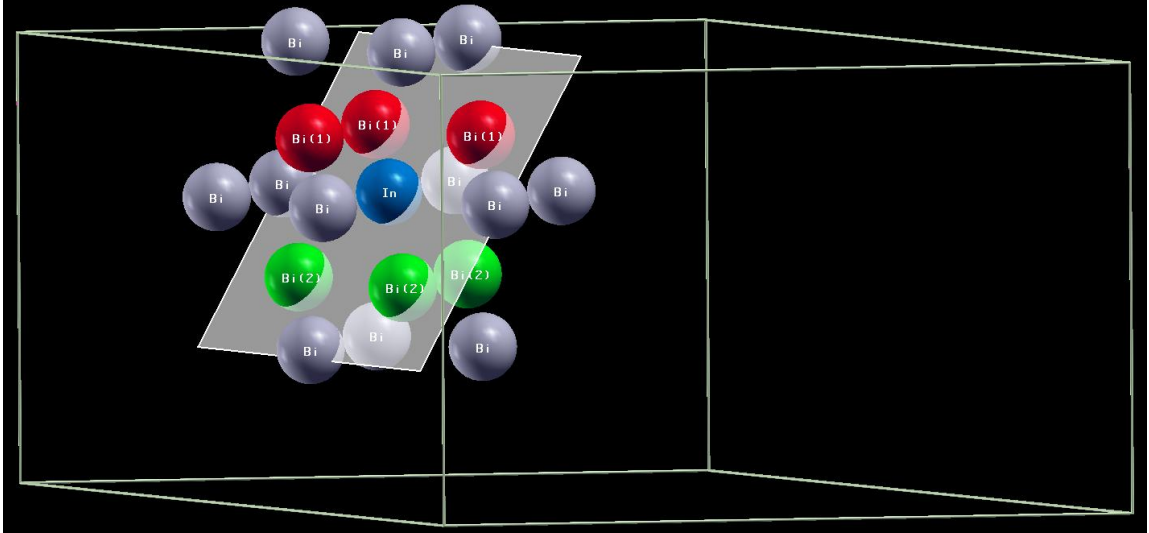

*Figure S3. Neighborhood of an In atom and the plane on which the charge density is projected in Fig. 3 of the main paper. Created using XCrySDen [9].*

1.4. Discussion of the doping efficiency of In in Bi

We found in III.1.1 of the main text that the doping efficiency of In is lower than that of substitutional In, which should release two holes in Bi. This indicates that not all In atoms substitute for Bi atoms, likely due to the presence of interstitial impurities in the Bi lattice that compensate the p-type doping action of substitutional In. Such behavior was previously identified for the case of Li-doped Bi,[10] which showed transport properties of a degenerately doped n-type semiconductor. A simple electron count would have identified substitutional Li as an acceptor in Bi. KKR-CPA calculations[10] showed that the strong n-type behavior of the Li-doped Bi samples can be explained if Li is an interstitial impurity, where it acts as a simple electron donor, rigidly moving the $E_F$ deep into the conduction band. Here, two types of defects, the presence of interstitial In and interstitial Bi, were considered in the KKR-CPA calculations on Bi:In. In the first case, we assume that some In atoms substitute for Bi, but the rest end up being at interstitial positions. In the second case, we assume that all In atoms are at substitutional positions, but push some of Bi atoms into interstitial sites.

The largest interstitial site, the 1a (0,0,0) site, lies between Bi atoms along the trigonal axis, and was selected for the calculations (inset in Fig. S4(a)). Fig. S4(a) shows the DOS of Bi with interstitial In atoms only. Interstitial In acts as an electron donor, moving the $E_F$ toward the conduction band. The combination of interstitial and substitutional In in the Fig. S4(b) shows that interstitial In compensates the acceptor behavior of the substitutional In. Here, the concentrations of the atoms in calculations were selected in order to result in the Bi:In atomic ratio as seen in the $Bi_{0.995}In_{0.005}$ substitutional case. This was done to mimic the experimental condition where only the Bi:In atomic ratio is controlled. Interstitial In gives one electron to the main valence block, and for small In concentrations ~ 0.1%, $E_F$ moves according to this number. For higher concentrations near the pseudo-gap, interstitial In acts more effectively, and the position of $E_F$ moves approximately as if it had the valence up to three for ~ 1%.

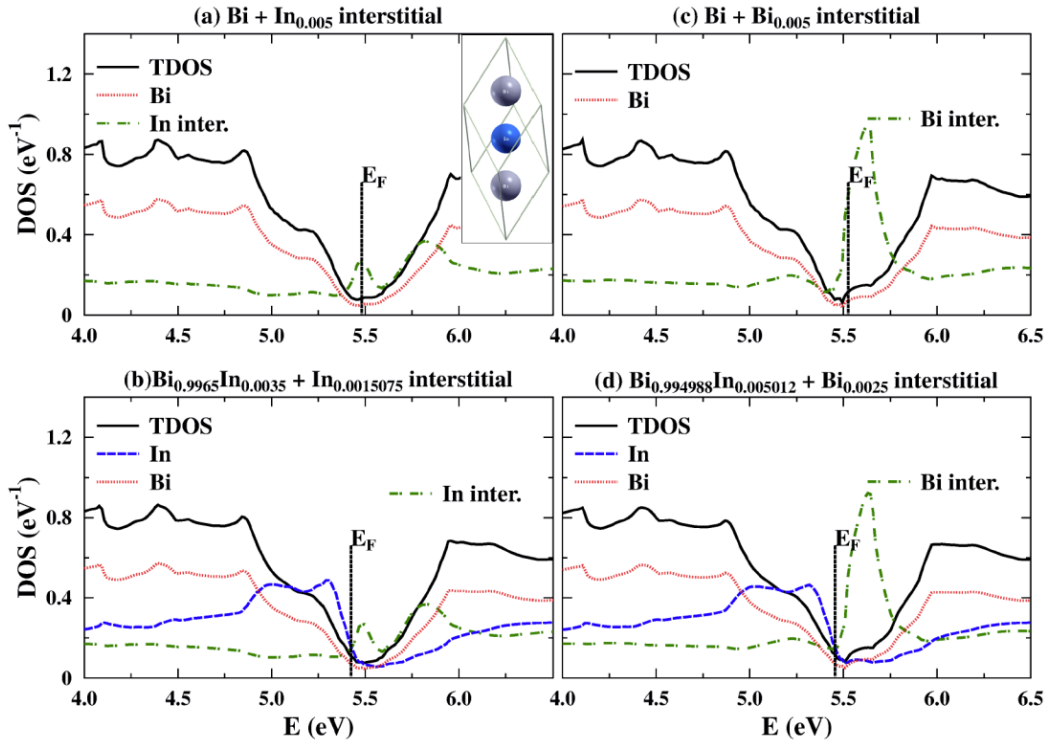

*Figure S4. KKR-CPA density of states (DOS) for Bi and $Bi_{1-\alpha}In_\alpha$ with defects: (a) interstitial In in undoped Bi, (b) interstitial In in $Bi_{1-\alpha}In_\alpha$, (c) interstitial Bi in undoped Bi, (d) and interstitial Bi in $Bi_{1-\alpha}In_\alpha$. The inset in (a) shows an In atom (blue ball) located at the interstitial site between Bi atoms (gray balls) in the rhombohedral unit cell of Bi. The random-looking concentrations are from the attempt to keep the atomic ratio close to the $Bi_{0.995}In_{0.005}$ case.*

A similar behavior is observed in the second case, that of interstitial Bi (Fig. S4(c) and (d)). Generally, interstitial Bi is a strong electron donor, with the valence number up to three, and in combination with the substitutional In, this pair behaves as an n-type impurity. Thus, to make the system p-type, the number of interstitial Bi should be at least two times smaller than the number of substitutional In.

In either case, the behavior of the Bi:In system with defects is not a simple rigid shift of $E_F$, which makes it difficult to define a precise valence number of the defects. Even so, a meaningful conclusion can be drawn from this defect study: either interstitial In or interstitial Bi may lead to the carrier compensation effects and explain the low efficiency observed for In doping in Bi.

1.5. Gallium-doped bismuth: theory

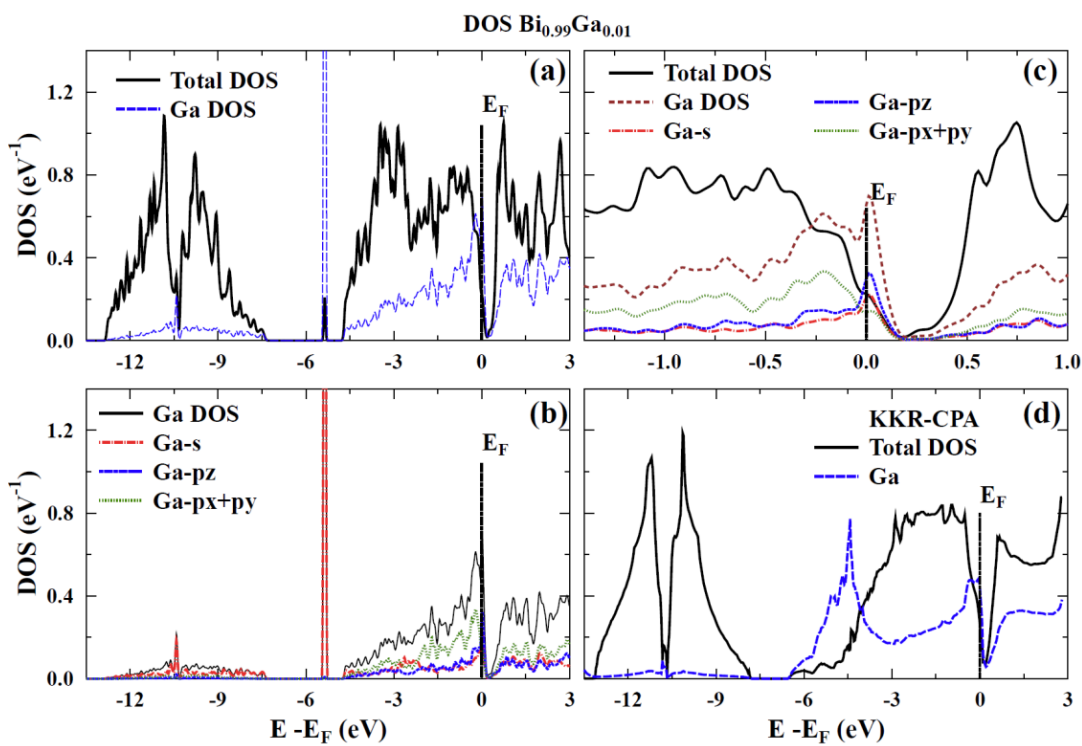

*Figure S5. Density of states (DOS) for $Bi_{0.99}Ga_{0.01}$. (a-c) FP-LAPW supercell results, (d) KKR-CPA result.*

A set of calculations as were performed on In-doped Bi were also applied to Ga-doped Bi, with the same technical details. Being a smaller atom than In, substitutional Ga creates a stronger relaxation effect when placed in the Bi matrix: the nearest neighbors move from 3.06 Å to 2.89 Å and the next nearest from 3.51 Å to 3.38 Å. Nevertheless, this difference does not alter the main features of the DOS of the doped system. Fig. S5(a) and (b) show the formation of the HDS DOS peak in the supercell study, with no formation of the resonant deep defect state (DDS) near $E_F$ (Fig. S5(c)). As in the case of Bi:In, the HDS peak is also present in the KKR-CPA results, but broadened and overlapped with the valence band DOS (Fig. S5(d)). The acceptor behavior of Ga is well seen in Fig. S6, where $E_F$ moves deeper into the valence band upon increasing the concentration of Ga atoms. Therefore, clear similarity between In and Ga doped Bi is observed, in agreement with the experimental observations shown later. This suggests that the p-type doping mechanism via creation of an HDS is not limited to the Bi:In case, and is likely to be applied to other dopants and host materials.

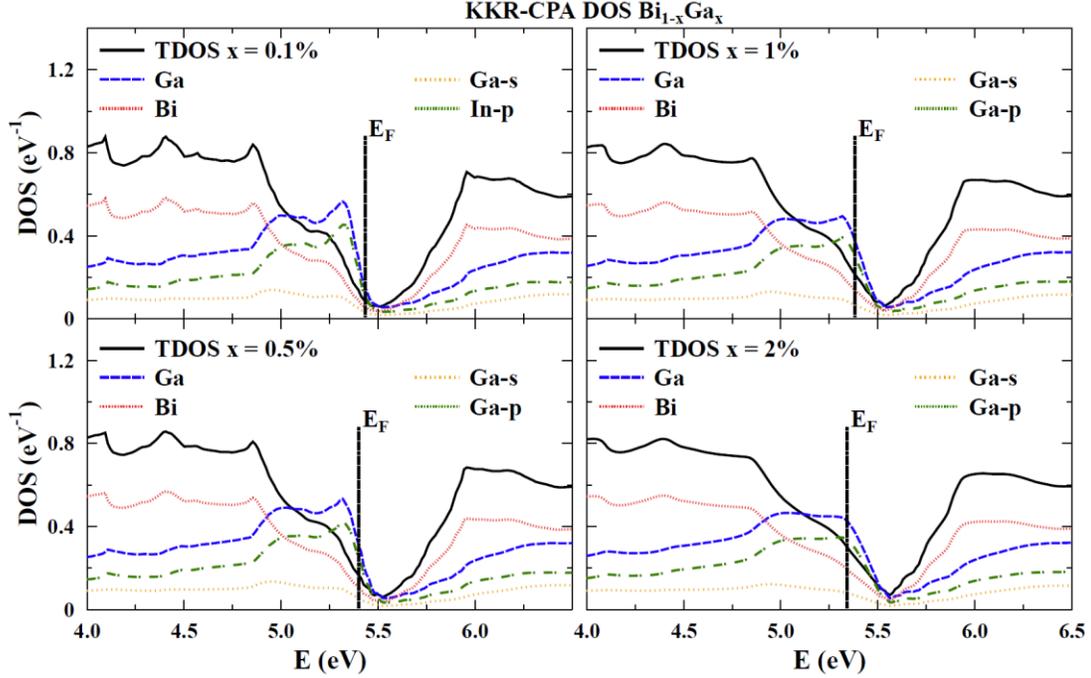

*Figure S6. Evolution of density of states (DOS) for the Bi:Ga from the KKR-CPA calculations. The Ga and Bi atomic contributions are given per atom, not multiplied by the concentrations.*

1.6. Tin-doped bismuth: KKR-CPA results

The KKR-CPA total and atom-decomposed DOS near $E_F$ for 0.1%, 0.5%, 1%, and 2% substitutional Sn in Bi are presented in Fig. S7, where contributions from Sn $s$- and $p$-states are also plotted. We again observe the confirmation of acceptor behavior of Sn and the absence of DDS formation (i.e. there are no partial DOS peaks near $E_F$).

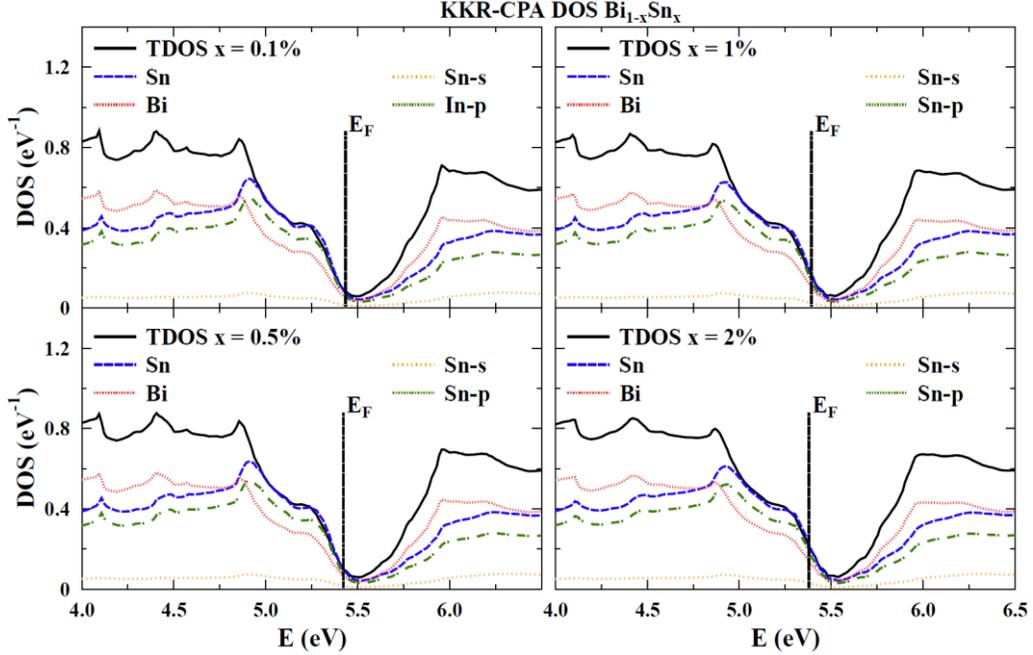

*Figure S7. Evolution of density of states (DOS) for the Bi:Sn from the KKR-CPA calculations. The Sn and Bi atomic contributions are given per atom, not multiplied by the concentrations. Modifications of DOS near $E_F$ are closer to rigid band like, compared to the In and Ga cases.*

## 2. Experiments

### 2.1. Indium-doped bismuth: experiments

#### 2.1.1. Equations for galvanomagnetic phenomena, single crystals, electrons and holes

Here we report the detailed analysis of the Hall measurement for Bi:In samples shown in Fig. 6(b) in the main text. In elemental Bi single crystals, electrons of density $N$ dominate the low-field Hall coefficient[11] when the conditions $\mu_x\mu_y B_z^2 \cong 1$ and $v^2 B_z^2 < 1$ hold simultaneously. Here, $\mu_x$ and $\mu_y$ are the electron mobilities taken for each electron ellipsoid along the $x$ (binary) and $y$ (bisectrix) axes, respectively [inset in Fig. 6(a)], and $v$ is the isotropic hole mobility in the $xy$-plane. In this regime, we observe a negative slope in $\rho_{xy}(B_z)$ for both Bi:In samples [inset in Fig. 6(b)]. In contrast, $\rho_{xy}(B_z)$ becomes nearly linear in $B_z$ with a positive slope at higher fields where $v^2 B_z^2 > 1$, which indicates that holes of density $P$ now dominate. The slope of each curve corresponds to Hall coefficient $R_H$ of each sample. In the low-field limit, $R_H$ yields the electron concentration, while in the high-field limit, $R_H$ reflects the excess hole concentration. In this case, $\lim_{B \to 0} R_H = -C/Nq$ and $\lim_{B \to \infty} R_H = C/(P-N)q$, where $q$ is the electron charge and $C$ is the Hall prefactor for the $\rho_{xy}(B_z)$ configuration, given by[12,13]:

$$C = 4\left(\mu_x\mu_y - \frac{P}{N}v^2\right)\left(\mu_x + \mu_y + 2\frac{P}{N}v\right)^{-2} \quad \text{(S1)}.$$

When $B_z \to 0$, Eq. (S1) can be reduced to $C = 4\mu_x\mu_y/(\mu_x + \mu_y)^2$. By inserting mobility values for pure Bi taken from Ref. [12] at 4.2 K, $C \cong 0.1$. Here, we assumed that the ratio $\mu_x/\mu_y$ is not affected by In doping, which is justified, since the ratio between the electron effective masses near the Fermi energy is not affected either. Additionally, it is observed that variation in $C$ from 4.2 K to 10 K is negligible.[12] Therefore, the same $C \cong 0.1$ is used for $T = 2$ K in Fig. 6(b) of the main text. When $B_z \to \infty$, $\rho_{xy}(B_z)$ becomes linear and thus, $R_H$ saturates, indicating that the material becomes degenerate. In degenerate semiconductors or semimetals with spherical constant energy surfaces $C = 1$ [Ref. 14]. $N$ and ($P$-$N$) for each In-doped sample can be calculated using the obtained $R_H$ and $C$, and these are the values reported in Table II of the main paper.

### 2.1.2. Equations for the thermopower, with two carrier types and two scattering mechanisms

A model is developed here to compute the total thermopower of a system with both electrons and holes, each of which is subject to two scattering mechanisms. As in the main text, the index $\phi$ refers to the scattering by acoustical phonons, while the index $r$ refers to the scattering due to the added impurity. When there are both electrons and holes, the total $S$ for each crystallographic direction can be expressed as

$$S = \frac{S_e \sigma_e + S_h \sigma_h}{\sigma_e + \sigma_h} \quad \text{(S2)},$$

where $S_e$ ($< 0$) and $S_h$ ($> 0$) are the partial thermopowers of electrons and holes, respectively. Moreover, in a similar way with the Matthiessen's rule, one can combine $S$ for different scattering mechanisms using the Gorter-Nordheim rule:

$$S = \frac{S_\phi/\mu_\phi + S_r/\mu_r}{1/\mu_\phi + 1/\mu_r} \quad \text{(S3)}.$$

From Eq. (S2) and (S3), an expression for the total $S$ of the In-doped samples is found for each direction:

$$S = \frac{Nq\left[\dfrac{S_{e\phi}/\mu_\phi + S_{er}/\mu_r}{(1/\mu_\phi + 1/\mu_r)^2}\right] + Pq\left[\dfrac{S_{h\phi}/\nu_\phi + S_{hr}/\nu_r}{(1/\nu_\phi + 1/\nu_r)^2}\right]}{Nq\left(\dfrac{\mu_\phi + \mu_r}{\mu_\phi \mu_r}\right)^{-1} + Pq\left(\dfrac{\nu_\phi + \nu_r}{\nu_\phi \nu_r}\right)^{-1}} \quad \text{(S4)}.$$

To evaluate the partial thermopowers, which are isotropic,[15] the variation of the $E_F$ with temperature can be computed from the known effective masses and temperature dependence of $N$ and $P$. Heremans et al.[16] introduced a pseudo-parabolic model for Bi which takes into account the non-parabolicity of the conduction band at the L-point, as well as the temperature dependence of electron effective masses. In the model, the relation between the Fermi energy, as measured from the band edge, $E_F$, and $N$ is given by

$$N = \frac{16\pi}{3h^3}(2\det\mathbf{m_e})^{1/2}\int_0^\infty \gamma(E)^{3/2}\left(-\frac{\partial f_0}{\partial E}\right)dE \qquad (S5).$$

Here, $\mathbf{m_e}$ is the band-edge mass tensor of electrons, whose determinant is the cube of the density of states mass, $\gamma(E) = E(1 + E/E_G)$ where $E_G$ is the gap in the energy spectrum, and $f_0$ is the Fermi distribution function.

While the model has been successfully applied to explain the behavior of Sn-doped Bi samples,[17] the authors found that it does not provide an adequate temperature dependence of $E_F$ at high temperature. The difficulty lies in the unknown temperature dependence of the heavy electron mass along the bisectrix direction. We observe that the pseudo-parabolic model works even at high temperatures when the temperature dependence of the electron effective mass is ignored while the non-parabolicity is kept.[16] Therefore, in our calculation, the mass determinant in Eq. (S5) was assumed to be temperature independent. The partial thermopower of electrons, a scalar, is given by the pseudo-parabolic model[16] as

$$S_e = -\frac{k_B}{q}\left[\frac{\left(\frac{5}{2}+\lambda\right)F_{3/2+\lambda}(\eta_F) + \left(\frac{7}{2}+\lambda\right)F_{5/2+\lambda}(\eta_F)/\eta_G}{\left(\frac{3}{2}+\lambda\right)F_{1/2+\lambda}(\eta_F) + \left(\frac{5}{2}+\lambda\right)F_{3/2+\lambda}(\eta_F)/\eta_G} - \eta_F\right] \qquad (S6),$$

where $\lambda$ is the scattering parameter, defined as the exponent of the energy dependence of the relaxation time $\tau \propto E^\lambda$, $F$ is the Fermi integral:

$$F_r = \int_0^\infty \frac{\eta^r}{1+\exp(\eta-\eta_F)}d\eta \qquad (S7),$$

and $\eta_F$ and $\eta_G$ denote $E_F/k_BT$ and $E_G/k_BT$, respectively. As regards the T-point holes, the dispersion can be effectively described by the parabolic model for which equations are obtained by setting $E_G \to \infty$ in Eq. (S5) and (S6):

$$P = \frac{16\pi}{3h^3}(2\det\mathbf{m_h})^{1/2}\int_0^\infty E^{3/2}\left(-\frac{\partial f_0}{\partial E}\right)dE \qquad (S8).$$

$$S_h = \frac{k_B}{q}\left[\frac{\left(\frac{5}{2}+\lambda\right)F_{3/2+\lambda}(\eta_F)}{\left(\frac{3}{2}+\lambda\right)F_{1/2+\lambda}(\eta_F)} - \eta_F\right] \qquad (S9).$$

With the known temperature dependence of $N$ and $P$, the temperature dependence of $E_F$ for electrons and holes can be found from Eq. (S5) and (S8). Those $E_F$'s are in turn substituted into Eq. (S6) and (S9) to yield the partial thermopowers for electrons and holes, respectively. For undoped Bi, $\lambda$ = -1/2 denoting the acoustical phonon scattering, while $\lambda$ = 0 is added for the In-doped sample to account for the effect from the energy-independent neutral impurity scattering. The resistivity provides evidence for temperature-independent scattering, which implies the energy-independent character of the scattering mechanism.

2.2. Gallium-doped bismuth: experiments

One polycrystalline sample of Bi, to which 0.5 at. % Ga was added, was prepared in a vacuum sealed ampoule. The ampoule was heated to 903 K and then rapidly quenched. This method usually does not result in a polycrystalline alloy with fine and randomly-oriented grains, because, upon solidification, the liquidus front progresses radially inward, which leads to a slight preferential crystallographic orientation of the grains in the final sample. This is difficult to characterize quantitatively, so that measurements of strongly anisotropic properties are sample-dependent and thus unreliable. We confine the results here to comparing the Ga-doped sample to a similarly-prepared undoped Bi sample, and to an analysis of the zero-field resistivity and the Hall effect, which reveals the density of the majority and minority carriers no matter what the crystal orientation is.[18] The possibility of Ga segregation was studied by DSC, but, contrary to the case of Bi:In, no Ga segregation was detected down to 10 ppm level. Therefore, we assume that most of 0.5 at. % Ga has been dissolved in the Bi matrix.

The zero-field $\rho(T)$ of the Bi:Ga sample resembles that of Bi:In showing a bump at 45 K, whereas the undoped polycrystalline Bi sample exhibits a metal-like behavior (Fig. S8(a)). The overall magnitude of $\rho(T)$ of Bi:Ga is larger compared to that of Bi:In, indicating enhanced alloy scattering, possibly because the difference in Pauling electronegativity is larger between Bi and Ga than between Bi and In. Fig. S8(b) shows that the Hall resistivity $\rho_H$ is a linear function of magnetic field ($B$) at the high field regime where $\bar{\nu}^2 B^2 > 1$, which suggests that holes, of density $P$, dominate. We use the notation $\bar{\nu}, \bar{\mu}$ to denote the mobility of holes and electrons, respectively, averaged over all crystallographic directions. On the other hand, in the low field limit where the conditions $\bar{\mu}^2 B^2 \cong 1$ and $\bar{\nu}^2 B^2 < 1$ hold simultaneously, we observe a negative slope in $\rho_H(B)$ (inset in Fig. S8(b)), indicating the presence of a small number ($N$) of high mobility minority electrons.

The method used by Issi[18] to characterize two-band conduction in undoped Bi polycrystals with $P=N$ is quite different from the method described in the main text (III.1.2) with Eqs. (1-2) for single crystals; we copy it here and extend it to the case where $P \neq N$. When two-band conduction is considered, the four parameters to be fitted are the electron density $N$ and mobility $\bar{\mu}$, and the hole density $P$ and mobility $\bar{\nu}$. The electrical conductivity ($\sigma = 1/\rho$) at zero field is the sum of the contribution of electrons and holes:

$$\sigma = \sigma_e + \sigma_h, \quad \sigma_e = Nq\bar{\mu}, \quad \sigma_e = Pq\bar{\nu} \qquad (S10).$$

The relative transverse (i.e. with the current flow perpendicular to $B$) magnetoresistance is fitted with two parameters $a$ and $b$ to a magnetic field dependence:

$$\frac{\rho(B) - \rho(B=0)}{\rho(B=0)} = \frac{aB^2}{1+bB^2} \qquad (S11),$$

where parameters $a$ and $b$ are given by:

$$a = \frac{1}{q} \frac{\sigma_e \sigma_h}{(\sigma_e + \sigma_h)^2} (\bar{\mu} + \bar{\nu})^2$$

$$b = \frac{1}{q^2} \frac{(\sigma_e \sigma_h)^2}{(\sigma_e + \sigma_h)^2} \frac{(N-P)^2}{N^2 P^2} \qquad (S12).$$

Here, the experimental value of parameter $b$ is only reliable in the high field limit, whereas $a$ can be fitted at intermediate with good accuracy. As was the case for single crystals, the expected saturation of $\rho(B)$ at high field is not always observed, which forces us to refrain from using $b$.

Therefore, a measurement of the magnetic field dependence of the Hall resistivity $\rho_H(B)$ is added (Fig. S8(b)) for the four parameter fit. Again, following Issi[18]:

$$\rho_H = \frac{B}{q} \frac{q^2(P\bar{v}^2 - N\bar{\mu}^2) + B^2 \frac{\sigma_e^2 \sigma_h^2}{q^2} \frac{P-N}{N^2 P^2}}{\sigma^2 + B^2 \frac{\sigma_e^2 \sigma_h^2}{q^2} \frac{(P-N)^2}{N^2 P^2}} \quad (S13),$$

which we can separate into a low-field Hall coefficient $R_{H0} \equiv \lim_{\bar{\mu}B, \bar{v}B \to 0} (\rho_H / B)$ and a high-field Hall coefficient $R_{H\infty} \equiv \lim_{\bar{\mu}B, \bar{v}B \to \infty} (\rho_H / B)$ given by:

$$R_{H0} = q(P\bar{v}^2 - N\bar{\mu}^2) / \sigma^2$$
$$R_{H\infty} = \frac{1}{q} \frac{1}{P-N} \quad (S14).$$

The final step is to use experimental values of $\rho$, $a$, $R_{H0}$, and $R_{H\infty}$ to derive values of $P$, $N$, $\bar{\mu}$, and $\bar{v}$. For Ga-doped polycrystalline Bi, we find $N \cong 9.4 \times 10^{15}$ cm$^{-3}$ and $P \cong 1.3 \times 10^{18}$ cm$^{-3}$ at 2 K, with mobility values of $\bar{\mu} \approx (5 \pm 1) \times 10^5$ cm$^2$V$^{-1}$s$^{-1}$ from 6 K to 60 K, and $\bar{v} \approx 1 \times 10^5$ cm$^2$V$^{-1}$s$^{-1}$ below 10 K, which decreases to $\bar{v} \approx 2.5 \times 10^4$ cm$^2$V$^{-1}$s$^{-1}$ at 60 K. The carrier concentrations are shown in Fig. S8(c) as a function of temperature. Above 60 K, the full model does not result in unique and accurate values for the properties of the minority electrons, but the excess hole concentration $(P - N)$ can still be obtained accurately just by using $R_{H\infty}$ in Eq. (S14). The data confirm that Ga, another group III element, also behaves as an acceptor in Bi, and its doping efficiency (i.e. the number of holes per Ga atom) is estimated to be higher than that of In. The similarity in the behavior of $\rho(T)$, and the high mobility of the electrons, which is temperature-independent below 60 K, reveals that Ga introduces the energy-independent neutral impurity scattering in Bi that In does.

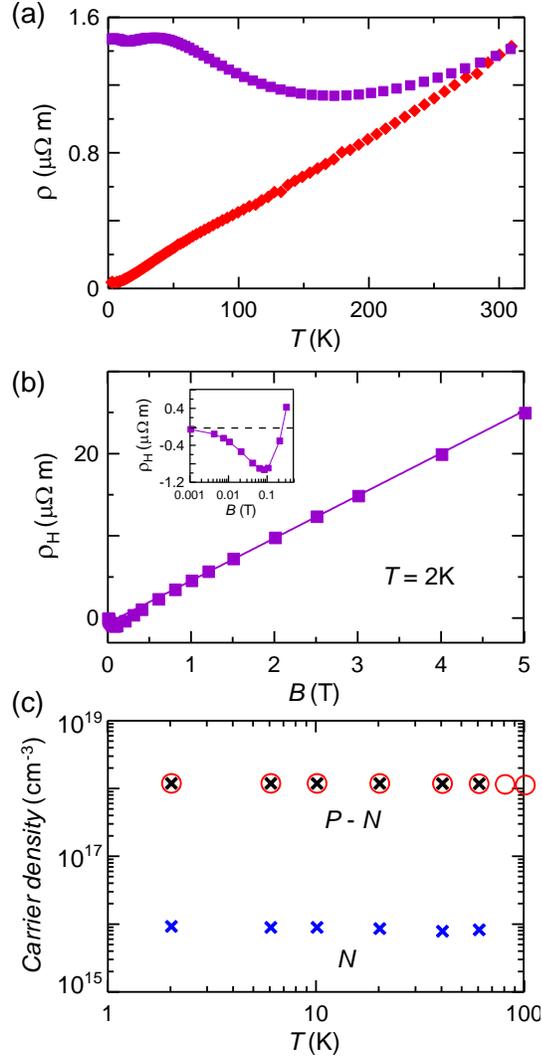

*Figure S8. (a) Temperature dependence of the electrical resistivity ρ of undoped Bi and $Bi_{99.5}Ga_{0.5}$ polycrystalline samples. The symbols are the experimental data: (red diamond) undoped Bi, (purple square) $Bi_{99.5}Ga_{0.5}$. (b) Hall resistivity $\rho_H(B)$ versus magnetic field for $Bi_{99.5}Ga_{0.5}$ sample measured at 2K. The points indicate the experimental data, while the lines are added to guide the eye. The inset contains magnification of $\rho_H(B)$ at low magnetic field, which shows the transition from the negative slope to the positive slope. (c) Temperature dependence of the electron (N) and excess hole (P - N) densities for $Bi_{99.5}Ga_{0.5}$ sample. The cross symbols are the data obtained by solving Eqs. (S10-S14), and the circle symbols indicate the excess hole density calculated from the slope of $\rho_H(B)$ at high magnetic field (i.e. $R_{H\infty}$ in Eq. (S14)).*